\documentclass[12pt,preprint]{aastex}

\usepackage{enumerate}

\nocite{*}

\newcommand{\gsim}{\hbox{\rlap{\lower.55ex\hbox{$\sim$}} \kern-.3em
\raise.4ex \hbox{$>$}}}
\newcommand{\lsim}{\hbox{\rlap{\lower.55ex\hbox{$\sim$}} \kern-.3em
\raise.4ex \hbox{$<$}}}

\slugcomment{Not to appear in Nonlearned J., 45.}

\shorttitle{1.5273~$\mu$m DIBs from APOGEE \emph{hot telluric} calibrators}
\shortauthors{Elyajouri et al.}

\begin{document}


\title{A catalog of 1.5273~$\mu$m diffuse interstellar bands based on APOGEE \emph{hot telluric} calibrators}


\author{M. Elyajouri,\altaffilmark{1}\email{meriem.el-yajouri@obspm.fr} A. Monreal-Ibero\altaffilmark{1},  Q. Remy
\altaffilmark{2} and R. Lallement\altaffilmark{1}}
\affil{1 GEPI Observatoire de Paris, PSL Research University, CNRS, Universit\'e Paris-Diderot, Sorbonne Paris Cit\'e\\}
\affil{2 Laboratoire AIM, IRFU/Service d'Astrophysique CEA/DSM
CNRS Universite Paris Diderot}



\altaffiltext{1}{5, Place Jules Janssen,
92195 Meudon, France}
\altaffiltext{2}{Bat. 709, CEA-Saclay, 91191
Gif-sur-Yvette Cedex, France}


\begin{abstract}
High resolution stellar spectroscopic surveys provide massive amounts of diffuse interstellar bands (DIBs) measurements. Data can be used to study the distribution of the DIB carriers and those environmental conditions that favor their formation. In parallel, recent studies have also proved that DIBs extracted from stellar spectra constitute new tools for building the 3D structure of the Galactic Interstellar Medium (ISM). The amount of details on the structure depends directly on the quantity of available lines of sight (LOS). Therefore there is a need to construct databases of high-quality DIB measurements as large as possible.
We aim at providing the community with a catalog of high-quality measurements of the 1.5273~$\mu$m DIB towards a large fraction of the \emph{Apache Point Observatory Galactic Evolution Experiment} (APOGEE) hot stars observed to correct for the telluric absorption and not used for ISM studies so far. This catalog would complement the extensive database recently extracted from the APOGEE observations and used for 3D ISM mapping.
We devised a method to fit the stellar continuum of the hot calibration stars and extracted the DIB from the normalized spectrum. Severe selection criteria based on the absorption characteristics are applied to the results. In particular limiting constraints on the DIB widths and Doppler shifts are deduced from the HI 21 cm measurements, following a new technique of decomposition of the emission spectra. 
From  $\sim$16\,000 available \emph{hot telluric} spectra we have extracted $\sim$ 6700 DIB measurements and their associated uncertainties. The statistical properties of the extracted absorptions are examined and our selection criteria are shown to provide a robust dataset. The resulting catalog contains the DIB total equivalent widths, central wavelengths and widths. We briefly illustrate its potential use for the stellar and interstellar communities.

\end{abstract}





\section{Introduction}

The spectra of stars viewed through one or several interstellar clouds display a plethora of relatively weak non-stellar absorption features of unknown origin, the so-called Diffuse Interstellar Bands \citep[DIBs, see][for a review]{Herbig95,Sarre06}. First DIBs were discovered around the early 20's by \citeauthor{Heger22} \citep[see][for a review of the history of the DIBs discovery]{McCall13} and  their interstellar origin was established in the 30's \citep{Merrill34,Merrill36}. Today, more than 400 optical DIBs {  (400-900 nm)} are known \citep[e.g.][]{jenniskens94,Hobbs09}. The first near-infrared (NIR) DIB was discovered by {  \citet{Joblin90}}. Currently, $\sim$25 {  NIR } DIBs have been identified {  \citep{Foing94,Geballe11,Cox14,Hamano15,Hamano16}}, including the strong band at 1.318~$\mu$m. NIR DIBs are particularly useful since they allow to make use of highly reddened target stars and explore the densest areas of the ISM. 
Most measured DIBs have  a Galactic origin. Still, they have also been detected in the Magellanic clouds, M\,31 and M\,33 {  \citep{Welty06,Cordiner08a,Cordiner08b,Ehrenfreund02,Cordiner11,vanLoon13} }and in a few line-of-sights towards e.g. starburst galaxies or in Type Ia supernovae spectra {  \citep{Sollerman05,Cox08,Heckman00,Phillips13}}. Recently a DIB gradient was established for the first time in a 160 Mpc-distant galaxy  \citep{MonrealIbero15}.

In spite of the number and ubiquity of DIBs, the carrier of most DIBs (i.e. the agents that originate these features) remains unidentified. 
Carbon is involved in most of the proposed candidates and among them we can find hydrocarbon chains \cite[e.g.][]{Maier04}, polycyclic aromatic hydrocarbons {   \citep[PAHs, e.g.][]{Vanderzwet85,Leger85,Crawford85,Salama96,Kokkin08}, and/or fullerenes \citep[][]{IglesiasGroth07,Sassara01}. See \cite{Cox11} for a recent review on the DIB-PAH hypothesis. } Today the situation is particularly promising in this regard, since the carrier for two DIBs {  at  9577 and 9632 \AA\ } was recently for the first time unequivocally identified {  with C$_{60}^+$ }  \citep{Campbell15}, confirming earlier results of \cite{Foing94}. {  C$_{60}^+$ was also detected in emission towards NGC7023 by \cite{Berne13}, } and C60 and C70 have also been identified in emission in young planetary nebulae \citep{Cami10}. Besides, first insights on the physical properties of the environments that are favorable or not for DIB formation are being obtained from high-quality absorption data  \citep{Cox06,Vos11,Cordiner13}. 

An interesting property of these features is that they are correlated with the amount of interstellar matter along the line of sight. Various correlations with neutral hydrogen, extinction
and interstellar Na\,I\,D and Ca\,H\&K lines have been established \citep[e.g.][]{Herbig93,Friedman11}. As a result, DIBs can be used to trace the structure of the ISM in the same way than others species, and they even offer certain advantages when used instead of (or in addition to) other tracers. For example, given their intrinsic weakness, they are ideal tracers in conditions where other features (e.g. Na\,I\,D) saturate, like very dense molecular clouds or regions seen through a large amount of extinction. Encouraged by this correlation between DIBs and IS matter, several teams have recently presented works that made use of the information provided by the different spectroscopic surveys to study the Galactic ISM structure and extinction in 2D or 3D by using the strength of different DIBs as a proxy \citep[e.g.][]{Yuan14,Kos14,Puspitarini15,Lan15,Baron15}. 

In the near-IR, the 1.5273~$\mu$m band (hereafter, we will also use 15273 \AA\  {  and refer to vacuum wavelengths})  falls in the spectral range observed by the \emph{Apache Point Observatory Galactic Evolution Experiment}  \citep[APOGEE,][]{Wilson10,Majewski12}, and this high resolution, high signal, massive survey  was used by \citet{Zasowski15}  to trace the 3D structure of the ISM at the large scale based on an unprecedented number of extracted NIR DIBs. Such pioneering work has demonstrated the potential of these methods and consequently that increasing the number of measurements will result in more accurate mapping.

In this article we present a work based on a fraction of the spectra of the most recent APOGEE's target sample made publicly available. The entire dataset comprises $\sim$160\,000 stars, most of them cool K and M giants ($T_{eff}\sim3500-5000$\,K). As part of the observing strategy and calibration plan of the survey, $\sim$17\,000 \emph{hot telluric} stars were also observed (see details in Sec. \ref{secdata}). By definition these are the bluest stars on a given pointing, and thus, most of the times early (i.e. O- to F-type) stars. As such, their almost featureless continua make them ideal targets to extract the information associated to the 1.5273~$\mu$m DIB. Our work has been focused on these calibration stars.
We have devised a method to fit their stellar continua and we present the resulting catalog of DIB measurements extracted from their spectra. This catalog complements the existing extensive catalog of \cite{Zasowski15} based on the APOGEE main, cool targets.

The paper is structured as follows:
Sect. \ref{secdata}  contains  the technical aspects regarding the data used in this work. 
Sect. \ref{secmethod} describes the methods used for continuum fitting and DIB parameter extraction, whereas  a detailed description of our criteria to validate a DIB detection and estimate uncertainties is included in Sect \ref{secvali}.
Sect. \ref{seccatalog} describes the resulting catalog and its validation. Sect. \ref{applications} illustrates some potential applications.
Conclusions are summarized in Sect. \ref{secconclu}.

\section{The data \label{secdata}}

\subsection{APOGEE overview}
The SDSS-III \emph{Apache Point Observatory Galactic Evolution Experiment} (APOGEE) is one of the four Sloan Digital Sky Survey III
\citep[SDSS-III,][]{Eisenstein11,Aihara11} experiments operated from 2011 to 2014. 
APOGEE uses a 300-fiber multi-object spectrograph working in the {  near infrared} ($H$-band, 1.51- 1.70~$\mu$m) at high spectral resolution \citep[R$\sim$22\,500,][]{Wilson10}. The spectrograph is attached via a fiber optic train to the SDSS 2.5-meter telescope at Apache Point Observatory \citep{Gunn06}. 
Since the effects of extinction at near-IR wavelengths are reduced, APOGEE has a strong potential to explore dust obscured regions of the Galaxy that are beyond the reach of optical surveys, in particular the inner Galaxy and bulge.
In a given observation, about 230 (out of 300) fibers are allocated for scientific targets selected from the source catalog the \emph{Two Micron All Sky Survey} \citep[2MASS, ][]{Skrutskie06}.
The $\sim$160\,000 selected scientific targets are distributed across all Galactic environments. Most of them are red giant stars but there are also special subsamples including stars with measured parameters and abundances from other spectroscopic studies, cluster members, etc. associated to several APOGEE ancillary science programs \citep[see ][for a detailed description of the different target classes]{Zasowski13}.
 
In addition to the science targets, a sample of hot stars, also called telluric standards stars (TSSs)  or \emph{hot telluric} calibrators are observed (35/300 fibers) to allow for the correction of the telluric absorption lines (see below). Finally, the remaining $\sim$35 fibers are allocated for observing blank sky positions to enable sky emission subtraction. 

\subsection{The APOGEE telluric standard stars}

The total number of TSSs in the APOGEE survey is nearly $\sim$ 17\,000.
Ideally, for a good telluric correction one would like to have bright O, B and A stars because of their featureless spectra. 
Therefore, to maximize the chances of observing stars of these types, the TSSs were selected as the bluest stars on a given plate having a magnitude in the range 5.5$\leq$H$\leq$11 and ensuring a uniform  distribution over the plate \citep[see][for details of TSS selection]{Zasowski13}.
Their spectra are used to adjust in an exquisite way the atmospheric transmission and its variability throughout the field, a transmission that is used to remove the telluric lines in all observed targets. 

{  It is important to note that the TSSs are corrected for telluric absorption lines in the same way as the science targets. In other words, they benefit from the correction tool they have themselves produced.} An interesting point about these stars is that, in addition to their use as tools to decontaminate the telluric absorption, they constitute a highly valuable dataset for ISM studies since their smooth continua are ideal to extract the information associated with {  DIBs. Here we focus on the strongest 1.5273~$\mu$m DIB. Work is in progress on the weaker absorption bands.}

The TSS can easily be identified in the APOGEE data products by means of the \texttt{APOGEE\_TARGET2} bitmask since all these stars are flagged as \texttt{APOGEE\_TELLURIC} (bit 9).

\subsection{Used information extracted from the APOGEE Archive}

This work is based on the products from the SDSS data release 12 \citep[DR12,][]{Alam15}. This release includes all the data taken between April 2011 and July 2014. The spectra can be downloaded from the Science Archive Server (SAS) as described on the data access page \footnote{\texttt{http://data.sdss3.org/sas/dr12/apogee/spectro/redux/\\r5/stars/l25\_6d/v603/}}. There are three different types of reduced spectra available \citep[see][]{Nidever15,Holtzman15}.
Here, we use the calibrated, well-sampled, and pseudo continuum-normalized combined 1D spectra. These are available as fits files with four separate extensions: the first one carries the information about the star and the derived stellar parameters, the second one contains the observed spectra, the third one saves the error pixels and the fourth one has a state-of-the-art modeled spectrum $S_\lambda$ adjusted to the observed one \citep{Garcia15}. The modeled spectra are interpolated from a grid of spectra, itself based on a large sample of model atmospheres \citep{Meszaros12}. 
These modeled spectra are optimized for stars at temperature $3\,800$~K$<T<5\,250$~K \citep{Ahn14}. The TSSs are \emph{a priori} hotter, and therefore as shown below, small adjustments to the provided modeled spectra were needed.

All the spectra are sampled at same rest wavelength pixel scale, with a constant dispersion in  $\log\lambda$, as 
\begin{equation}
\log \lambda_{i}= 4.179 + 6 \times 10^{-6} \: i
\label{pixel}
\end{equation}
with 8575 total pixels (i = 0 to 8574), giving a rest wavelength range of 15\,100.8 to 16\,999.8~\AA~\citep{Nidever15}.

We also use the table that summarizes all the parameters derived from the APOGEE spectra (for DR12, allStar-v603.fits), including for each observed individual star its mean barycentric radial velocity, the standard deviation of the mean velocity, the mean ASPCAP parameters and abundances as derived from the combined spectra, and a compilation of ancillary targeting data \citep[see][for a full description of the data in these files]{Nidever15,Holtzman15}.

\section{Data analysis \label{secmethod}}

The aim of this work is analyzing the series of  $\sim$16\,000  DR12 TSS spectra {  than can be downloaded from the APOGEE site} to determine the presence (or absence) of the DIB at 
15273~$\AA$ and measuring its equivalent width in case of positive detection.
Given the volume of data, this needs to be done with as less as possible human intervention.
For that, we took as starting point the methodology presented by \citet{Puspitarini13} for optical spectra, and adapted it to the peculiarities of APOGEE spectra.  At the end, we had a fully automated DIB extraction method able to measure DIB equivalent widths (\emph{EWs}) without any user interaction during the fit. The code was developed using the IGOR PRO environment\footnote{\texttt{http://www.wavemetrics.com}}. Note that the reduced pseudo-normalized spectra of DR12 are corrected from telluric lines contamination. Thus, no telluric correction is needed.
In the following, we give the details of our fitting technique, we explain how we extracted the EW of the DIB and how we estimated their associated uncertainties.

\subsection{Profile fitting} 

We fit each TSS spectrum to a model made out of the product of several components as follows:

\begin{equation}
M_\lambda= [S_\lambda]^\alpha \times DIB[\sigma,\lambda,D] \times (1+[A] \times \lambda)
\label{fit}
\end{equation}

with:\\

\begin{figure}[ht]
   \centering
   \includegraphics[width=0.9\columnwidth]{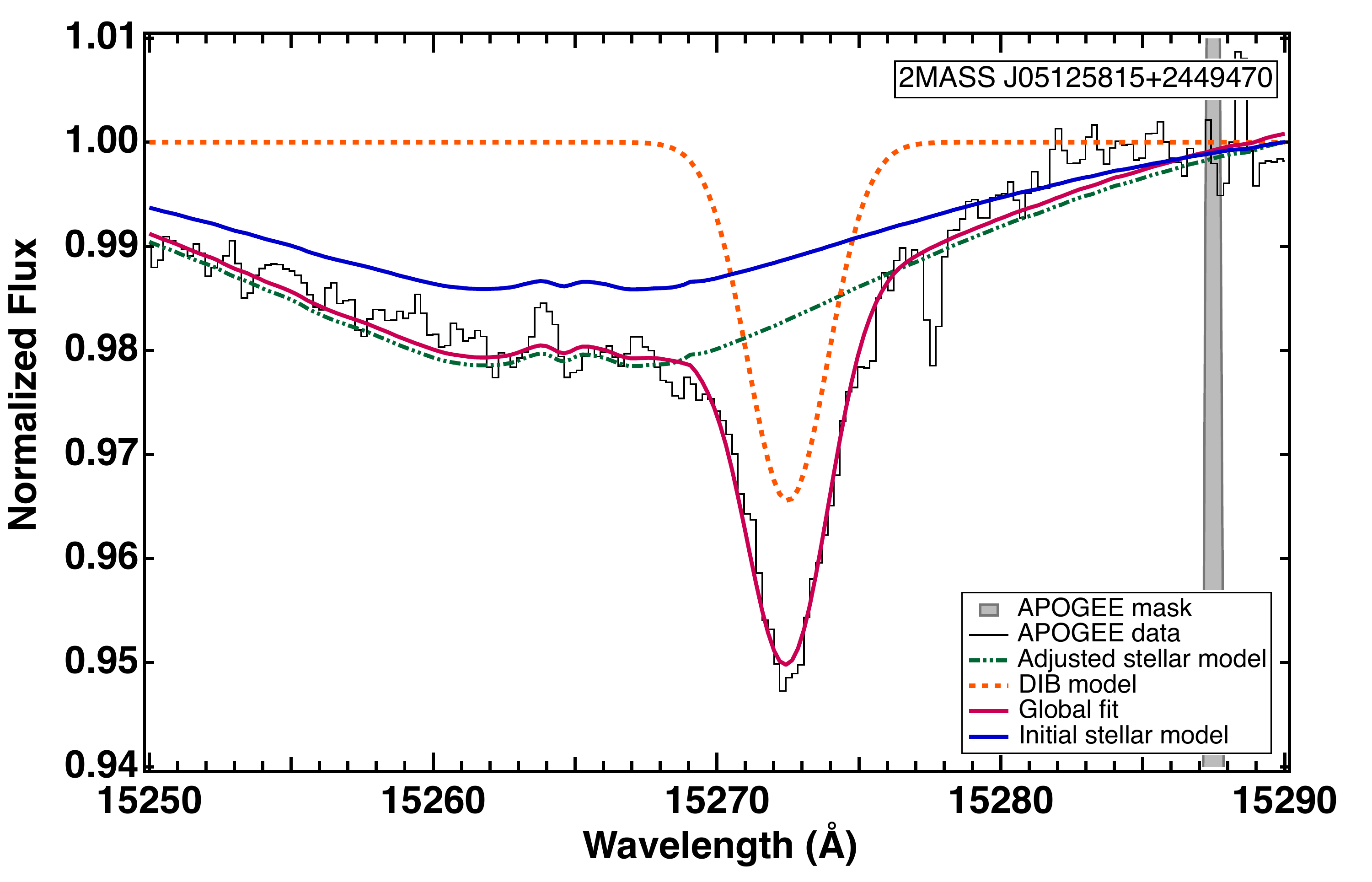}
      \caption{Example illustrating our fitting method. Data as provided by the APOGEE survey are shown in black while the final  stellar+DIB modeled spectrum appears in magenta. The initial stellar model  provided by APOGEE and the model obtained after application of the scaling factor $\alpha$ are shown in solid blue and dash-dot green lines respectively. The DIB absorption alone is represented with a dashed orange line. A grey band in the red part of the spectrum shows an example of masked region.}
         \label{Figfittingmethod}
\end{figure}

$\bullet$ $[S_\lambda]^\alpha$, \emph{an adjusted stellar spectrum}: $S_\lambda$ is the initial stellar model provided by the APOGEE project. As we mentioned above, this model is not optimized for the TSSs that are hotter than the main targets. Therefore, even if the velocity of each target star was accurately determined and the global shape of the spectrum was most of the times adequate, the depth of the stellar atmospheric lines was not properly estimated (see Figure \ref{Figfittingmethod}). We included a scaling factor ($\alpha$) in order to take this into account and to adjust the model depth to the data.\\  

$\bullet$ $DIB[\sigma,\lambda_c,D]$, \emph{the DIB profile}: It was modeled as a Gaussian function with three free parameters associated to its width ($\sigma$), central wavelength ($\lambda_c$) and depth ($D$).\\

$\bullet$ $(1+[A] \times \lambda)$, \emph{a local continuum}: this was a simple 1-degree polynomial introduced to model as closely as possible the continuum around the DIB.\\
 
We selected a pre-defined spectral range for the fit restricted to the vicinity of the DIB [$15\,260-15\,290]\AA$ to determine the above coefficients $\alpha$, $\sigma$, $\lambda_c$, $D$ and $A$. Note that this range was large enough to ensure an adequate sampling of the neighboring Bracket 19-4 stellar line at 1.52647 $\mu$m. Errors provided by APOGEE were used to mask those spectral ranges affected by artifacts due to imperfect sky emission correction or other sources of uncertainty. We use the Levenberg-Marquardt algorithm implemented in IGOR PRO to compute the coefficients that minimize the chi-square $\chi^2$.
An example illustrating our fitting procedure is shown in Figure \ref{Figfittingmethod}.

\subsection{DIB equivalent width and error estimates} 

In the following, we explain how we extracted the DIB equivalent width from our fits and how we estimated the uncertainties associated to each of the parameters presented in the catalog.\\
$\bullet$ \emph{Computation of the DIB equivalent width:} Since none of the previous or present observations reveals any asymmetry of the 15273 \AA\ DIB, and for our sightlines the linear regime prevails, we assume that the actual DIB absorption has a shape that is similar to the mono-gaussian model. We also neglect the small departures from a Gaussian that result from multiple cloud superimposition that is small compared to other sources of uncertainties. Subsequently, assuming that the continuum $I_0(\lambda)$ is reasonably fitted, the DIB equivalent width ($EW$) is  the area of the fitted Gaussian and can be simply analytically derived from its parameters as the product of the model DIB's depth $D$ by the coefficient $\sigma$: 

\begin{equation}
EW= \int_{DIB} \frac{I_0 - I_\lambda}{I_0}\,d\lambda = \sqrt{2 \pi} \: D\:  \sigma 
\label{moneq}
\end{equation}
where $I(\lambda)$ is the observed spectrum.

$\bullet$ \emph{Computation of standard deviations in residuals:}
We calculated the standard deviation of the fit residuals, \emph{data - model}, in a region close to the DIB (region A=[$\lambda_{c}$ - 10, $\lambda_{c}$ + 10]  \AA).
In addition to this, we performed a second fit using the same function as the one presented in equation \ref{fit}  but over the whole spectral range covered by APOGEE. 
{  As a matter of fact, we found that the first and third terms of Equation 2 are good enough to represent continuum and Bracket lines in the entire APOGEE spectral region. For this fit the DIB term in the equation has the same restrictions as in the « local» fit. The DIB is fitted to ensure an optimum placement of the continuum in the DIB region and subsequently everywhere. As a matter of fact, omitting the DIB could bias the continuum in its spectral region and react on the whole fit.} The residuals from this additional fit were used to obtain two additional measurements of the standard deviation. The first one was calculated over a \emph{clean} (i.e.  free from DIB absorption, strong stellar lines and telluric residuals) spectral range (region B= [15\,892-15\,959]  \AA). The second one (region C= [15\,200-15\,250]  \AA) corresponds to a region relatively close to the DIB {  (and also free of DIB)} and potentially contaminated by stellar residuals.

Hereafter, we will call the standard deviation $R$  for all the three regions, $R_{A}$, $R_{B}$ and $R_{C}$.
These were used for the selection criteria discussed in Section \ref{secvali}. Also, the standard deviation  $R_{A}$ in the narrow spectral range around the DIB is used below to estimate the uncertainty on the DIB equivalent width.\\

$\bullet$  \emph{Uncertainties on the DIB central wavelength and width:}
The IGOR algorithm uses data and fitted model to evaluate the \emph{data$-$model} standard deviation in the spectral range used for the fit, within the implicit assumption that the fitting function is ideally adapted to the data and the data are characterized by random Gaussian noise. If there are telluric line residuals or any other artefacts that amplify the signal fluctuations, the estimated standard deviation is correspondingly increased and treated as random noise.  Based on the standard deviation and the mutual influence of the parameters the algorithm estimates one sigma errors on each parameter. We use these error estimates on both the central wavelength and the width.\\

$\bullet$ \emph{Uncertainty on the DIB equivalent width:}
For this quantity we estimated the error in a very conservative way in order to take into account the departures from Gaussian noise and the ideal fitting function, as well as potential subsequent misplacements of the continuum level. As a matter of fact in a number of cases telluric absorption or sky emission correction residuals are present and for a number of targets the stellar line shapes are not perfectly fitted. This may result in slight displacements of the fitted continuum that do not impact the line center and impact the width in a negligible way but may impact the depth. For the most conservative estimate of the error we used the standard deviation of region A, coupled to the maximum DIB width deduced for each sightline from HI 21 cm data (see Sect. \ref{secvali}).
The maximum error was estimated as the equivalent width of a Gaussian with the maximum width of the H\,I~21~cm line and a depth equal to $R_A$:
\begin{equation}
err(EW)=  \sqrt{2 \pi} \: R_{A}\:  \sigma_ {HI\,21cm}\
\label{eqdibhi21}
\end{equation}
We checked that this conservative error is appropriate to all types of spectra and data quality, including those cases where there are clear residuals linked to stellar or telluric lines. {  Figure \label{fig9add} shows the estimated error as a function of the equivalent width for all targets of the catalog. It can be seen from the figure that due to our method based on HI, for the same EW the error is significantly larger in directions where the ISM radial velocity range is very large, mainly here at Galactic longitudes smaller than 70 deg. This confirms that our quoted values are very conservative.}
\section{Validation criteria  \label{secvali}}
\subsection{Selection of fitted absorptions} 
We applied a series of tests to the fitted Gaussian absorptions based on the fitted parameters, the signal quality in the DIB region and elsewhere, and finally the LOS direction and the corresponding Galactic gas. The series of tests is schematically represented in Figure \ref{flowchart}.\\

$\bullet$ \emph{First global test}: the first selective test eliminates those absorptions that are deeper than 10\% of the continuum (arrow (a) in Figure \ref{flowchart}). Those depths are extremely unlikely for Galactic DIBs and especially for the TSSs that are nearby. Indeed, we checked a posteriori that all those cases correspond to spurious detections (stellar lines for the coolest or most metallic objects, unusually strong telluric residuals, etc). We also eliminated simultaneously absorptions centered outside the spectral range [15268-15280] \AA. In the red, the corresponding large Doppler shifts of $\geq$+137km~s$^{-1}$ are unrealistic for the nearby ISM. In the blue, the test excludes a tiny fraction of fits with unrealistically high negative Doppler shifts caused by a stellar line (see the histogram in Figure \ref{velocityhist}).\\
\\
$\bullet$ \emph{Test on the absorption depth}: as a second test for the selected targets (the arrow (b)),  we compared the depth of the fitted absorption line with the standard deviation on the residuals. This comparison was done in a one- or two-step process. We first compare the absorption depth with the maximum of the three standard deviations $R_{A}$, $R_{B}$ and $R_{C}$ described in section 3.2. If it is larger (arrow (d)), then start directly the tests on the width. If the depth is smaller that this maximum, i.e if the spectrum is globally too noisy to detect the DIB (the arrow (c)) or the DIB is very shallow, we also compared the absorption depth with the local standard deviation $R_{A}$ only. This second step allows to detect those spectra that are noisy in those spectral ranges far from the DIB  but that  have a good enough signal to noise in the DIB neighborhood to ensure a detection, i.e. the use of the maximum of the three $R_{A}$, $R_{B}$ and $R_{C}$ deviations is too restrictive. Absorptions that pass successfully the test (arrow (e)) are then examined for their widths in the same way than the previous ones (arrow(d)) but are flagged differently. The difference in the flags serves to evaluate the quality of the spectrum and identify those targets for which models have a shape incompatible with the measurements outside the DIB region. They will be visually inspected one by one. Absorptions who still appear too shallow (arrow (g)) are examined in a different way.\\

$\bullet$ \emph{Criterion on the width in case of shallow absorptions:} as a third test for those absorptions that appear small with respect to the noise, both locally and in the three spectral regions (i.e. from arrow (g)), we selected the DIBs with a width, $\sigma$, within predefined limits. The lower limit of 0.7 \AA\ is deduced from the results of  \cite{Zasowski15} and in particular from the $\sigma$ histogram of the whole APOGEE dataset shown in their Figure 6. This histogram shows that a very abrupt decrease occurs at $\simeq$ 1 \AA\, implying that the  intrinsic DIB width maybe as small as this value. Taking into account our typical uncertainties on the DIB width, err($\sigma$) $\simeq$0.35, we finally imposed a lower limit on the width of 0.7 \AA . This criterion has been checked a posteriori to eliminate spurious detections associated to bad pixels or telluric residuals, and simultaneously to keep actual detections of narrow DIBs. A typical example of such narrow absorptions that pass successfully the test  is shown in Figure \ref{dibsmall}. It can be seen from the figure that it looks convincingly like an actual DIB. The upper limit of 1.0 \AA\ is chosen for the following reasons: for those absorptions that are very shallow (arrow (g)), this maximum value eliminates flat and elongated features of uncertain origin. Conversely, those very shallow absorptions that are reasonably wide pass the test and are falling in the {\it upper limit} category (yellow box in Figure \ref{flowchart}). For those absorptions that are deeper and correspond to arrow (f), this maximum value will result in definitely rejecting those features that are broader than 1.5$\sigma_{HI21cm}$ (see below). \\

$\bullet$ \emph{Limits on the width $\sigma$ based on HI 21cm emission spectra}:
For all absorptions that exceed the noise level, either globally or locally (arrows (d) and (e)), a subsequent test on the DIB width is performed based on radio 21cm HI spectra. As a matter of fact the velocity range for some fitted absorptions is clearly not compatible with the velocity distribution in the Galaxy. Such unrealistic absorption widths are caused by an improper modeling of the stellar lines, residuals of telluric lines, or a locally increased level of noise. In order to eliminate these spurious or biased detections from the catalog we added the following constraint on the DIB width based on the radial velocities of the atomic hydrogen gas. 
Indeed as the HI is the most common and widespread form of gas in the galaxy, its extent in velocity should encompass the velocity extent of the DIB, as verified in previous DIB measurements \citep[see e.g. ][]{Puspitarini15,Zasowski15}.\\
We estimated the maximal radial velocity width of the DIB by constructing a \emph{virtual}, composite DIB based on the HI spectra. To do so we used the results of the HI line decomposition method presented in \cite{2015A&A...582A..31A}. We briefly recall the principle of this decomposition.
The first step is to detect the significant lines in the brightness temperature $T_{B}(v)$ spectra of the LAB Survey \citep{2005A&A...440..775K}. We measured the dispersion in temperature, T$_{rms}$, outside the bands with significant emission. In order to limit the number of false detections caused by strong noise fluctuation we clipped the data to zeros except if  $T_{B}(v)>T_{rms}$ in 3 channels (3 $km~s^{-1}$).
Then, using the 5-point-Lagrangian differentiation twice, we computed the curvature $d^2T_{B}/dv^2$ in each channel. Lines are detected as negative minima in $d^2T_{B}/dv^2$. False detections triggered by the edges of the clipped regions were eliminated.
Then each spectrum is fitted by a sum of pseudo-Voigt functions, one for each detected line. The pseudo-Voigt function is defined as
\begin{equation}
\rm{PV}= \eta L + (1-\eta) G \label{pv}
\end{equation}
with $0<\eta<1$ the shape factor, L a Lorentzian, and G a Gaussian respectively defined as :
\begin{equation}
L = \frac{h}{1+\left(\frac{v-v_{0}}{\sigma}\right)^2} \label{L}
\end{equation}
and,
\begin{equation}
G =  h \: {\rm exp}\left( \frac{v-v_{0}}{\sigma} \right)^2   \label{g}
\end{equation}
where $v_{0}$ is the position of the line, $\sigma$ the width and $h$  the height parameter.\\
In order to take into account the detection uncertainties, the position of the line $v_0$ is allowed to vary in a $\pm$ 4 $km~s^{-1}$ range around the initial position. For each detected line the free parameters $\eta$, $h$, $\sigma$, and $v_0$ are fitted by means of a $\chi^2$ minimization.\\
For each LOS the computed HI decomposition is used to 
construct a \emph{virtual} DIB, the one that would be measured in the spectrum 
of a star located beyond all the HI clouds, i.e. at the periphery of the 
Galactic HI disk. To do so, each HI component is associated to a DIB 
that has a width $\sigma_{0}$ = 1.4 \AA\ and a radial velocity equal to 
the one of the HI line. The chosen width $\sigma_{0}$ corresponds to an upper limit on the intrinsic DIB width that can be derived from the Figure 6 of \cite{Zasowski15}. To convert the HI line into a DIB equivalent width we use the following relationships:

\begin{enumerate}[i)]  
\item The HI column density is calculated in the optically thin limit: 
\begin{equation}
N_{HI} = 1.823 \times 10^{18} \int T_{b}(v) dv
\label{eq1}
\end{equation}
where $T_{b}(v)$ is the 21 cm line brightness temperature at radial velocity v.

\item The HI column density is transformed into  reddening {  using the $N_{HI}-E(B-V)$ empirical relation from \cite{Gudennavar12} {  and the average $R_v=A_v/E(B-V)$ value of =3.1:}} 
\begin{equation}
A_v=  3.1 \times 1.6\: \times10^{-22}\times N_{HI} \
\label{eq2}
\end{equation}

and finally 
\item The extinction is converted into a DIB equivalent width using the relation derived by  \cite{Zasowski15}: 
\begin{equation}
EW_{DIB}=  0.102 \times A_v^{1.01} (\rm \AA)\
\label{eq2}
\end{equation}
\end{enumerate}

The depth of each individual DIB is computed from the EW and the width. 
The final \emph{virtual} DIB is the product of those individual DIBs and as such it is 
broadened to match the velocity distribution of the HI spectrum for the entire path through the Galaxy. 
Figure \ref{dibvirtual} illustrates the construction of the \emph{virtual} DIB for 
a typical LOS with 6 HI clouds. This \emph{virtual} DIB is fitted to a Gaussian and the resulting width 
$\sigma_{HI21cm}$ is used to impose a limit on the APOGEE absorption 
width. Such a limit is ensuring the compatibility between the measured 
DIB width and the velocity spread seen in HI. Absorptions that do not 
pass the test (i) as shown in Figure \ref{flowchart} enter test (f). The broad absorptions will be discarded as discussed 
above and the narrow lines are reconsidered and classified as shown in Figure \ref{flowchart}.\\

\subsection{Special cases \label{special}} 
We applied target selection criteria and considered separately Be Stars and stars from the field 4529 because they can yield difficulties in fitting. To do so we used the \cite{Chojnowski15} catalog of Be stars and the information contained in the APOGEE tables. All these spectra have been the subject of one by one visual inspection and only those where the DIB was clearly unambiguously and accurately detected were kept.{   Figure \ref{illustration} } contains an example of a Be star spectrum and a spectrum from a star in plate 4259.

The different possible outcomes after the selection criteria described in section 4.1 and 4.2 have been flagged from 1 to 7. Figure \ref{illustration} presents one representative example for each category.

\section{The catalog  \label{seccatalog}}

A fraction of the catalog of all detected DIBs is presented in Tab. \ref{catalog} while a summary of its statistical properties appears in Tab. \ref{tabstatis}.
For the total of 6\,716 detections or upper limits the catalog lists the 2MASS identification, then the equivalent width, central wavelength and DIB Gaussian width and associated errors. The flag allowing to identify the selection criteria for each target is also added. Finally, we included for completeness the estimated equivalent width for the \emph{virtual DIB} as calculated from the emission in the H\,I~21 cm line.
The various categories (1) Upper limit (2) Stars from plate 4529 (3) Be Stars (4) Narrow recovered DIBs (5) Narrow DIBs (6) Recovered (7) Detected comprise 362, 11, 52, 122, 473, 572 and 5124 TSS targets respectively. {  Figure \ref{fig10add}  shows the location of all TSSs on a sky coordinates plot, distinguishing those that are included in the catalog. It can be seen that at Galactic latitudes smaller than $\simeq$  15 degrees all APOGEE fields contain TSSs with detected DIBs, while at higher latitude some fields have no targets included in this catalog, due to the weakness of the IS column.} \\

\begin{table}[ht]
\caption{Summary with statistics of the selected sample}             
\label{tabstatis}      
\centering                          
\begin{tabular}{c c c c}        
\hline\hline                 
     & $\lambda_c$ & $\sigma$ & $EW$ \\    
     & (\AA) & (\AA)& (m\AA) \\    
\hline                        
Mean & 15272.42 & 1.50 &  78\\
Std Dev &     1.13 & 0.40 &  49\\
Median & 15272.43 & 1.47 &  69\\
\hline                                   
\end{tabular}
\end{table}

The comparison  between the velocity distributions of the original sample and the one included in this catalog is presented in Figure \ref{velocityhist}.
Both of them peak at $\sim-20$~km~s$^{-1}$. However, while the final distribution resembles to a Gaussian distribution, the original one is quite irregular, with several secondary peaks. In particular those at $\sim-260$ and $-130$ ~km~s$^{-1}$ correspond to very strong stellar absorption features that have been successfully eliminated. Note that the final sample presents a relatively small (i.e $\lsim30$) number of detections with high velocities (i.e. $\sim100$~km~s$^{-1}$). After visual inspection of the corresponding spectra, we did not see any evidence for a spurious detection. Therefore we kept those detections in the catalog. They will be the subject of further work. We cannot exclude a similar high velocity tail in the blue due to the  threshold  imposed  to remove spurious identifications due to stellar lines.\\
Figure \ref{sigmahist} displays the distribution of the DIB widths. Only a very small fraction of selected DIBs  have sigma above 3 \AA. In the same way, Figure \ref{ewdibhist} shows that there are very few DIBs ($\sim$167) with equivalent width $>200$~m\AA. This is not unexpected since TSS are relatively nearby stars, much closer than the APOGEE targets full sample.
Finally, we show in Figure \ref{dibvsh21} a comparison between the APOGEE measured $EWs$ and the $EWs$ of the \emph{virtual} DIBs described in section \ref{secvali} that are proportional to the total hydrogen column from the LAB Survey \citep{2005A&A...440..775K}, converted into an equivalent width according to equation  \ref{eqdibhi21}.
The figure shows that, within the error bars, there is an envelope to all the data points that corresponds roughly to a one-to-one relation.
This supports the robustness of our method and the choice of the conversion factor. Data points close to the envelope very likely correspond to target stars located beyond all the \textsc{H\,i} clouds seen in emission, while points well below the envelope correspond to targets that are in front of at least one cloud. 
   

\begin{deluxetable}{cccccccccc}
\tabletypesize{\scriptsize}
\tablecaption{Sample stars with 15\,273 \AA\ DIB detection. \label{catalog}}
\tablewidth{0pt}
\tablehead{
\colhead{2MASS ID} & \colhead{GLON} & \colhead{GLAT} & \colhead{$\lambda_c\pm err(\lambda)$} & \colhead{$\sigma\pm err(\sigma)$} &
\colhead{$EW \pm err({EW})$} & \colhead{Flag} & \colhead{$EW_{H\,I~21 cm}$}\\
}
\startdata
$00001687+5903034$ & $116.390$ & $-03.169$ & $15274.0\pm0.2$ & $1.1\pm0.2$ & $041\pm018$ & 7&219\\
$00001877+5938132$ & $116.500$ & $-02.595$ & $15272.3\pm0.1$ & $1.4\pm0.1$ & $075\pm019$ & 7&231\\
$00005414+5522241$ & $115.740$ & $-06.791$ & $15271.1\pm0.3$ & $1.9\pm0.3$ & $058\pm018$ & 7&109\\
$00010331+1549325$ & $105.570$ & $-45.340$ & $15272.0\pm0.4$ & $0.8\pm0.5$ & $010\pm021$ & 1&017\\
$00010512+5651581$ & $116.060$ & $-05.332$ & $15271.9\pm0.4$ & $1.4\pm0.4$ & $027\pm014$ & 6&148\\
$00011569+6314329$ & $117.320$ & $+00.919$ & $15274.5\pm0.1$ & $1.5\pm0.2$ & $097\pm020$ & 7&313\\
$00013391+5904490$ & $116.550$ & $-03.172$ & $15272.2\pm0.1$ & $0.9\pm0.1$ & $042\pm022$ & 5&219\\
$00013919+7005554$ & $118.690$ & $+07.639$ & $15272.8\pm0.2$ & $1.6\pm0.2$ & $062\pm011$ & 7&187\\
$00022269+6254032$ & $117.380$ & $+00.559$ & $15273.5\pm0.1$ & $2.0\pm0.2$ & $128\pm025$ & 7&279\\
$00023939+7109185$ & $118.970$ & $+08.659$ & $15271.7\pm0.3$ & $1.7\pm0.3$ & $053\pm010$ & 7&138\\
$00024298+5748303$ & $116.460$ & $-04.450$ & $15272.4\pm0.2$ & $1.2\pm0.2$ & $033\pm016$ & 7&190\\
$00024671+1605384$ & $106.240$ & $-45.194$ & $15272.8\pm0.2$ & $0.7\pm0.2$ & $016\pm012$ & 5&017\\
$00024931+5709106$ & $116.350$ & $-05.096$ & $15272.7\pm0.3$ & $1.7\pm0.3$ & $049\pm013$ & 6&158\\
$00024931+5709106$ & $116.350$ & $-05.096$ & $15272.7\pm0.2$ & $1.3\pm0.2$ & $058\pm017$ & 7&158\\
$00025231+5841372$ & $116.650$ & $-03.584$ & $15272.9\pm0.2$ & $1.8\pm0.2$ & $069\pm017$ & 7&223\\
$00025606+7014169$ & $118.820$ & $+07.754$ & $15273.2\pm0.3$ & $1.2\pm0.3$ & $029\pm013$ & 7&164\\
$00032397+7310282$ & $119.420$ & $+10.631$ & $15268.4\pm0.5$ & $0.8\pm0.6$ & $008\pm016$ & 1&103\\
$00042334+5626552$ & $116.430$ & $-05.828$ & $15273.0\pm0.2$ & $0.8\pm0.2$ & $024\pm022$ & 4&141\\
$00044189+6939114$ & $118.860$ & $+07.152$ & $15274.4\pm0.3$ & $0.9\pm0.3$ & $021\pm013$ & 4&216\\
$00044627+0132128$ & $099.672$ & $-59.211$ & $15271.4\pm0.6$ & $0.9\pm0.6$ & $008\pm009$ & 1&013\\
$00045177+5933503$ & $117.060$ & $-02.775$ & $15272.5\pm0.4$ & $2.2\pm0.4$ & $066\pm027$ & 7&230\\
$00045306+7048114$ & $119.090$ & $+08.280$ & $15273.6\pm0.8$ & $1.4\pm0.8$ & $014\pm007$ & 6&138\\
$00050083+6349330$ & $117.840$ & $+01.414$ & $15271.9\pm0.1$ & $2.0\pm0.1$ & $175\pm030$ & 7&313\\
$00051293+5811530$ & $116.860$ & $-04.127$ & $15272.7\pm0.1$ & $1.2\pm0.1$ & $060\pm013$ & 7&223\\
$00051873+5758464$ & $116.830$ & $-04.345$ & $15274.0\pm0.2$ & $1.6\pm0.2$ & $070\pm024$ & 7&193\\
$00052394+6347207$ & $117.870$ & $+01.370$ & $15271.9\pm0.1$ & $1.7\pm0.1$ & $135\pm019$ & 7&313\\
$00052888+5717190$ & $116.730$ & $-05.028$ & $15272.0\pm0.3$ & $1.2\pm0.3$ & $026\pm014$ & 7&158\\
$00052965+5722141$ & $116.750$ & $-04.948$ & $15272.3\pm0.3$ & $1.4\pm0.3$ & $043\pm024$ & 7&158\\
$00053673+7530446$ & $120.010$ & $+12.901$ & $15272.2\pm0.2$ & $1.7\pm0.2$ & $082\pm023$ & 7&110\\
$00054538+6234237$ & $117.700$ & $+00.167$ & $15274.7\pm0.1$ & $1.4\pm0.1$ & $133\pm029$ & 7&293\\
$00060064+7001540$ & $119.040$ & $+07.504$ & $15271.3\pm0.3$ & $1.4\pm0.3$ & $036\pm011$ & 7&193\\
$00060945+6412287$ & $118.030$ & $+01.768$ & $15272.1\pm0.1$ & $1.7\pm0.1$ & $130\pm023$ & 7&335\\
$00061415+0127198$ & $100.290$ & $-59.414$ & $15271.8\pm0.2$ & $0.8\pm0.2$ & $023\pm016$ & 5&014\\
$00063033+5719168$ & $116.870$ & $-05.021$ & $15272.8\pm0.2$ & $0.8\pm0.2$ & $021\pm015$ & 5&172\\
$00064418+7541165$ & $120.110$ & $+13.062$ & $15271.0\pm0.4$ & $2.3\pm0.4$ & $071\pm026$ & 6&110\\
$00064478+0025574$ & $099.822$ & $-60.418$ & $15271.6\pm0.2$ & $0.7\pm0.2$ & $024\pm017$ & 4&019\\
$00064971+5737102$ & $116.970$ & $-04.734$ & $15272.7\pm0.4$ & $2.1\pm0.5$ & $048\pm016$ & 7&193\\
$00065832+5726544$ & $116.960$ & $-04.906$ & $15272.2\pm0.4$ & $1.8\pm0.4$ & $040\pm017$ & 7&172\\
$00071354+5804306$ & $117.100$ & $-04.295$ & $15272.8\pm0.2$ & $1.2\pm0.2$ & $041\pm023$ & 7&193\\
$00072008+6241381$ & $117.900$ & $+00.254$ & $15272.9\pm0.1$ & $1.4\pm0.1$ & $139\pm033$ & 7&296\\
$00074256+6318026$ & $118.040$ & $+00.845$ & $15271.3\pm0.1$ & $1.4\pm0.1$ & $120\pm022$ & 7&304\\
$00074657+5722435$ & $117.050$ & $-04.993$ & $15272.0\pm0.2$ & $1.4\pm0.3$ & $043\pm014$ & 7&172\\
$00074863+7040220$ & $119.310$ & $+08.109$ & $15271.7\pm0.6$ & $0.9\pm0.6$ & $009\pm010$ & 4&167\\
$00080292+7332356$ & $119.820$ & $+10.934$ & $15271.8\pm0.2$ & $0.9\pm0.2$ & $025\pm022$ & 3&099\\
$00081304+6315383$ & $118.090$ & $+00.796$ & $15273.9\pm0.1$ & $1.3\pm0.1$ & $085\pm023$ & 7&304\\
$00085840-0015231$ & $100.410$ & $-61.257$ & $15273.3\pm2.5$ & $0.9\pm2.7$ & $002\pm014$ & 1&016\\
$00090905+5737124$ & $117.270$ & $-04.786$ & $15271.8\pm0.2$ & $0.9\pm0.2$ & $036\pm015$ & 5&177\\
$00091817+7022314$ & $119.380$ & $+07.795$ & $15272.6\pm0.2$ & $1.2\pm0.2$ & $038\pm011$ & 7&167\\
$00092728+7447408$ & $120.130$ & $+12.152$ & $15271.4\pm0.2$ & $0.9\pm0.2$ & $038\pm012$ & 5&108\\
$00094423+0117063$ & $101.810$ & $-59.865$ & $15272.6\pm0.4$ & $1.7\pm0.5$ & $033\pm014$ & 7&013\\
\enddata
\tablecomments{Table \ref{catalog} is published in its entirety in the 
electronic edition of the {\it Astrophysical Journal}.  A portion is 
shown here for guidance regarding its form and content.
}
\end{deluxetable}


\section{Potential applications \label{applications}}

In the following, we illustrate the potential of this catalog by means of two examples: as a proxy for the extinction and as tool for ISM tomography.

The 15\,273 \AA\ DIB is correlated with the extinction over at least three orders of magnitudes \citep{Zasowski15}. Thus, it can be used as an independent approach to estimate it  or as an initial guess in absence of better information.
We crossmatched the present TSS catalog with the compilation of reddening measurements provided by \citet{Lallement14} used to build local 3D maps of the ISM.  Figure \ref{dib_vs_ebv} shows the comparison between the APOGEE 1.5273~$\mu$m DIB  equivalent width and the reddening for the 221 stars in common. There is a clear positive correlation showing that the fitted absorptions contain valuable information on the intervening ISM.
The figure also presents the linear fit derived by \citet{Zasowski15} for the APOGEE cool stars.
Our data points cluster around this relationship in solid agreement with the results obtained for the late-type stars. We note however that the number of \textit{outliers} with very weak DIBs is significantly higher than the number of \textit{outliers} with strong DIBs. A possible explanation to this trend could be that UV radiation field of the early-type stars acts on the intervening matter. This influence of the star temperature has already been noticed in the past \citep{Raimond12, Vos11}. 

Likewise, the catalog can be crossmatched with spectra available for the targets. Being relatively bright many of them have been the subject of specific investigations, allowing comparisons with other DIBs and atomic or molecular lines. We have found at least 100 targets that have been observed at very high resolution and are available from public archives. 

Finally, as mentioned in the introduction, our main interest in this (or similar) catalogs is the potential for use in ISM tomography. Compared with the extended catalog of  \citet{Zasowski15} that allows mapping the Galactic scales, the TSSs are more appropriate for the nearby ISM.  As a proof-of-concept we explored this use in the Taurus-Perseus region. This is illustrated in Figure \ref{plancktaurus} that shows the dust optical thickness measured by Planck in this area \citep{PlanckXI}. Superimposed on the map are the TSS targets color-coded according to the detection (or non-detection) of a DIB with an equivalent width higher than 50 m\AA. The strength of the DIB is always small in regions with negligible or small dust emission while the contrary is seen towards opaque clouds.
We selected two regions, marked as A and B in Figure \ref{plancktaurus}, to explore the use of this catalog to locate the molecular clouds along the line of sight. For that, we used the subset of targets in these areas with Hipparcos parallax \citep{Perryman97,Vanleeuwen07}.
Figure \ref{tworegions} shows a comparison between the detection (or absence) of a DIB towards a given target and the distance to it.  For both cases there is a clear cut-off in distance between this two categories. This allows to bracket the distance to the main absorbing cloud: targets without detection are in front of the clouds while those with detection are well beyond. 
For our particular cases the main clouds  are at about 150 pc, in agreement with the detailed tomography of \citet{Schlafly14}.  This illustrates the potential of this catalog to build detailed 3D maps of the ISM once the more accurate and numerous parallaxes provided by Gaia will become available.

 \section{Summary and perspectives \label{secconclu}}

We have analyzed the series of SDSS/APOGEE NIR spectra of the calibration stars used to decontaminate data from telluric absorption lines. Those targets are generally blue stars and the APOGEE \emph{Stellar Parameters and Chemical Abundances Pipeline} (ASPCAP, \citet{Garcia15}) stellar models cannot be used straight away. We have used them as a starting point and allowed the model broad stellar features to vary in depth to reproduce the observed continuum. In the region of the 1.5273~$\mu$m DIB this adjustment has proven to be good enough to allow the extraction of the DIB equivalent width and central radial velocity through Gaussian fitting of the DIB absorption.
Careful and severe examinations of the DIB parameters, the continuum shape and the quality  of the adjustment were done. This result in a conservative selection of reliable DIB parameters. In particular all DIB candidates that contradict radial velocity limiting values based on the HI 21 cm spectra have been excluded. A total of 6\,716 lines of sight are selected and the results are presented in a catalog that will be available from CDS, Strasbourg. The statistical properties of the resulting DIB database show that our selection criteria are fully appropriate. 

Since most of the calibration stars are nearby objects (say, within the first kpc), such a DIB catalog can be used to improve 3D maps of the nearby ISM and to complement maps at larger scale. We have compared our DIB detections in the Taurus-Perseus region with the distribution of dust as traced by Planck, finding a good agreement between the strength/existence of the DIBs and the location of the clouds in the plane of the sky. Likewise, we showed how used jointly with existing parallax (i.e. distances) of the stars they can be used to assign a distance to the clouds. A step forward would be the application to the DIBs of inversion methods to recover the 3D structure, as already attempted based on optical absorption lines and extinction \citep[e.g.][]{Vergely10,Lallement14, Sale14}. A strong advantage of NIR DIBs is the potential use of highly reddened target stars and the exploration of dense IS clouds. Moreover, nowadays instrumentation is becoming more and more NIR-IR oriented and large datasets in this spectral range are foreseen. 3D maps of the Galactic ISM are tools of wide use and  their quality is expected to increase considerably in future once the Gaia distances will be available.  Independently, DIB-based maps can be compared with dust and gas 3D distributions, providing new diagnostics of the conditions of their formation, ionization and destruction. They can also help interpreting small-structure of ISM clouds and of UV radiation impact on DIBs revealed by high spatial or temporal resolution spectroscopic data \citep{Cordiner13,Smith13}.

\acknowledgments

R.L., A.M.-I. and Q.R. acknowledge support from Agence Nationale de la Recherche through the STILISM project (ANR-12-BS05-0016-02). M.ElJ. acknowledges funding from the Region Ile-de-France through the DIM-ACAV project.\\
  Funding for the Sloan Digital Sky Survey IV has been provided by
the Alfred P. Sloan Foundation, the U.S. Department of Energy Office of
Science, and the Participating Institutions. SDSS-IV acknowledges
support and resources from the Center for High-Performance Computing at
the University of Utah. The SDSS web site is www.sdss.org.

SDSS-IV is managed by the Astrophysical Research Consortium for the 
Participating Institutions of the SDSS Collaboration including the 
Brazilian Participation Group, the Carnegie Institution for Science, 
Carnegie Mellon University, the Chilean Participation Group, the French Participation Group, Harvard-Smithsonian Center for Astrophysics, 
Instituto de Astrof\'isica de Canarias, The Johns Hopkins University, 
Kavli Institute for the Physics and Mathematics of the Universe (IPMU) / 
University of Tokyo, Lawrence Berkeley National Laboratory, 
Leibniz Institut f\"ur Astrophysik Potsdam (AIP),  
Max-Planck-Institut f\"ur Astronomie (MPIA Heidelberg), 
Max-Planck-Institut f\"ur Astrophysik (MPA Garching), 
Max-Planck-Institut f\"ur Extraterrestrische Physik (MPE), 
National Astronomical Observatory of China, New Mexico State University, 
New York University, University of Notre Dame, 
Observat\'ario Nacional / MCTI, The Ohio State University, 
Pennsylvania State University, Shanghai Astronomical Observatory, 
United Kingdom Participation Group,
Universidad Nacional Aut\'onoma de M\'exico, University of Arizona, 
University of Colorado Boulder, University of Oxford, University of Portsmouth, 
University of Utah, University of Virginia, University of Washington, University of Wisconsin, 
Vanderbilt University, and Yale University.    
    




\bibliographystyle{apj}
\bibliography{mybib_apj}

\appendix
\section{Various categories of selected absorptions}

\begin{figure}[ht]
   \centering
   \includegraphics[width=0.9\columnwidth, bb=0 0 610 540, clip=]{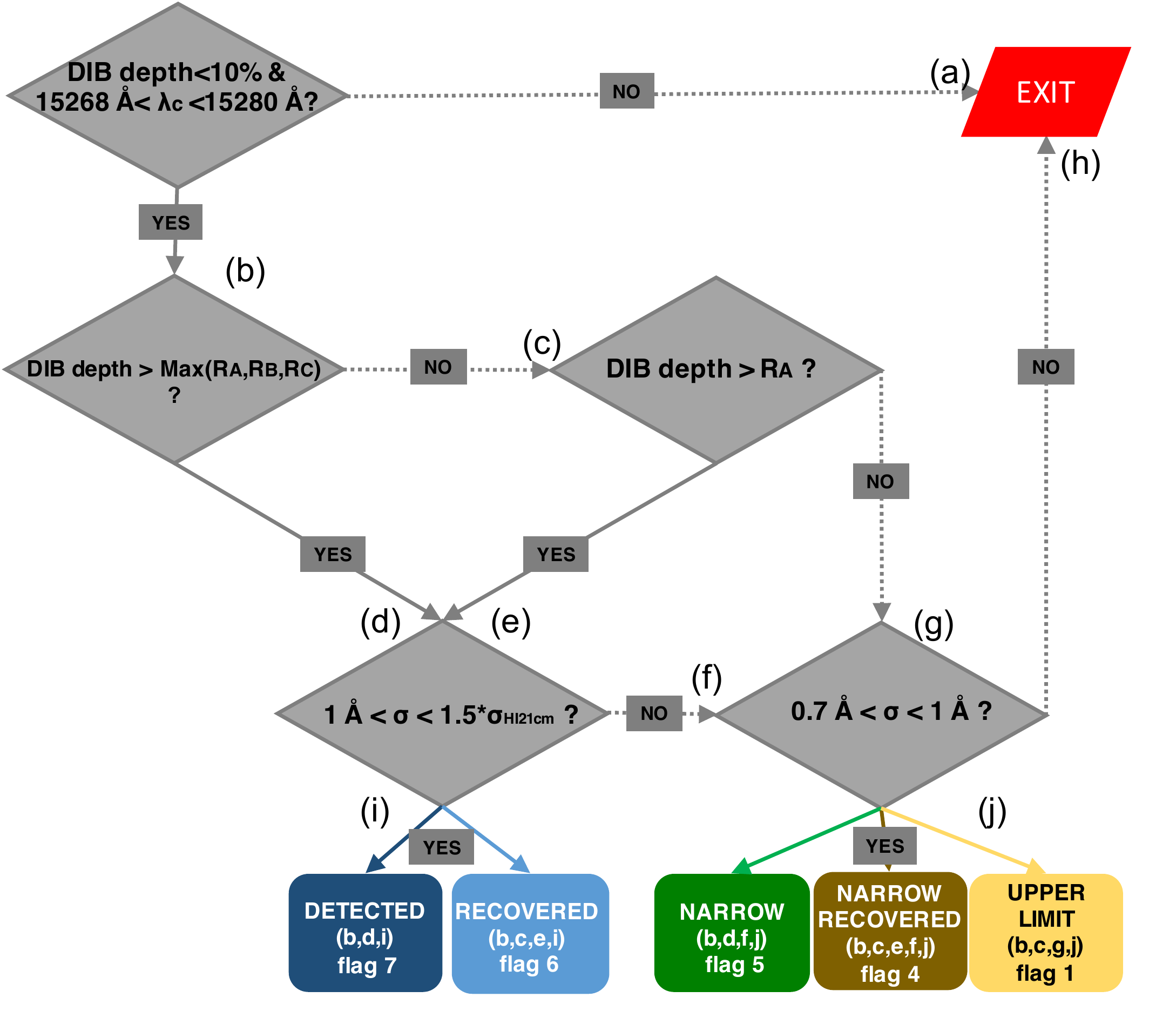}
      \caption{Flowchart compiling our decision criteria to create our catalog. Any spectrum able to reach the bottom part of the flow has a  positive DIB detection. The path followed by a given spectrum to be classified has been indicated in the corresponding box with small letters. {  Flags numbers in each box at bottom are detailed in section \ref{seccatalog} and included in the catalog.}}
         \label{flowchart}
   \end{figure}

\begin{figure}[H]
   \centering
   \includegraphics[width=0.9\columnwidth]{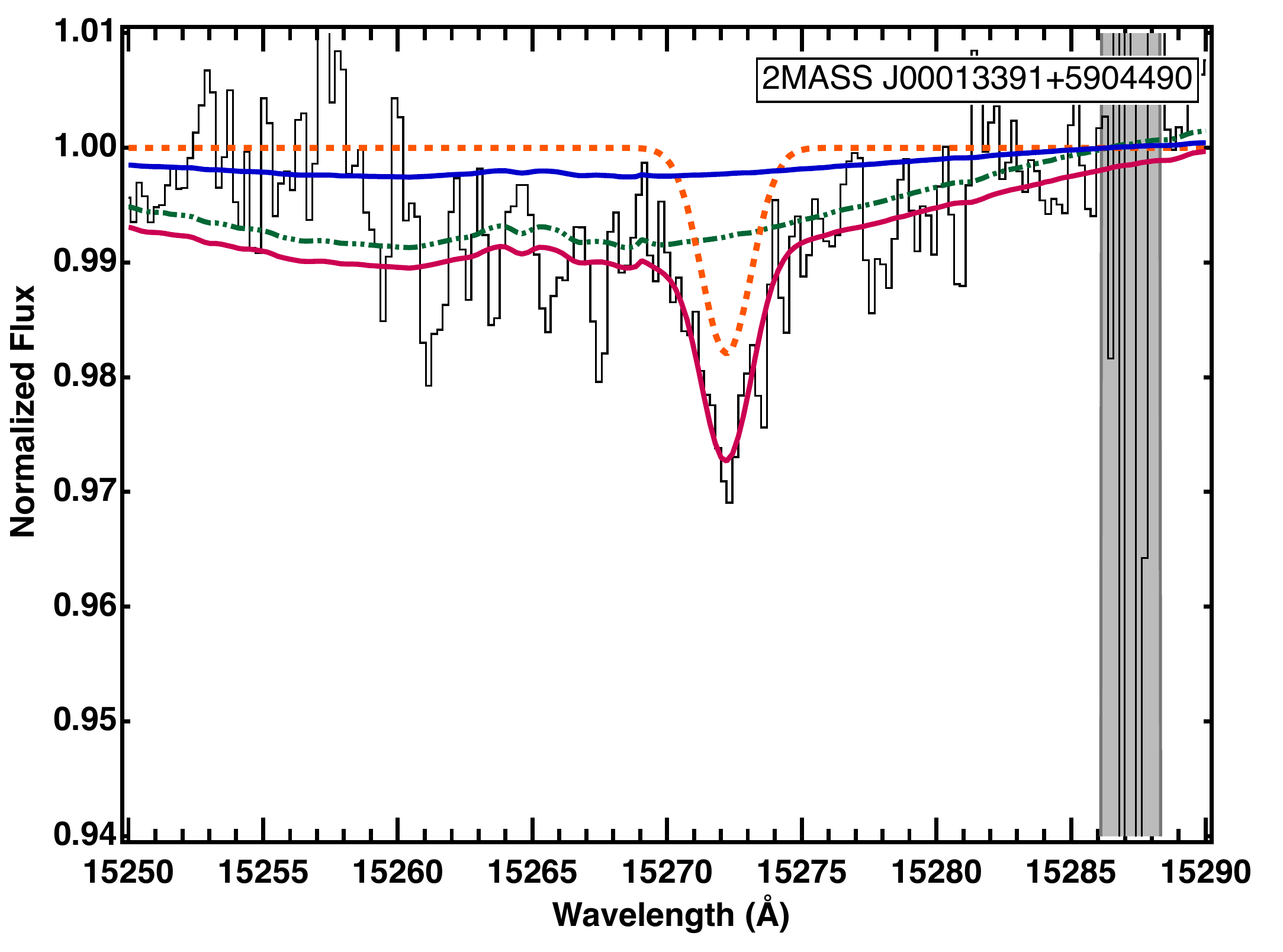}
      \caption{Example of detection of narrow DIB (0.7~\AA$<\sigma<$1\AA): 2MASS J00013391+5904490 with $\sigma= 0.9 \pm0.1$ $\AA$, $\lambda_{c}=15272.2 \pm 0.1$ $\AA$ and $EW=0.041\pm 0.021$ $\AA$. Color and symbol code is as in Figure \ref{Figfittingmethod}. }
         \label{dibsmall}
   \end{figure}

\begin{figure}[ht]
   \centering
   \includegraphics[width=0.9\columnwidth]{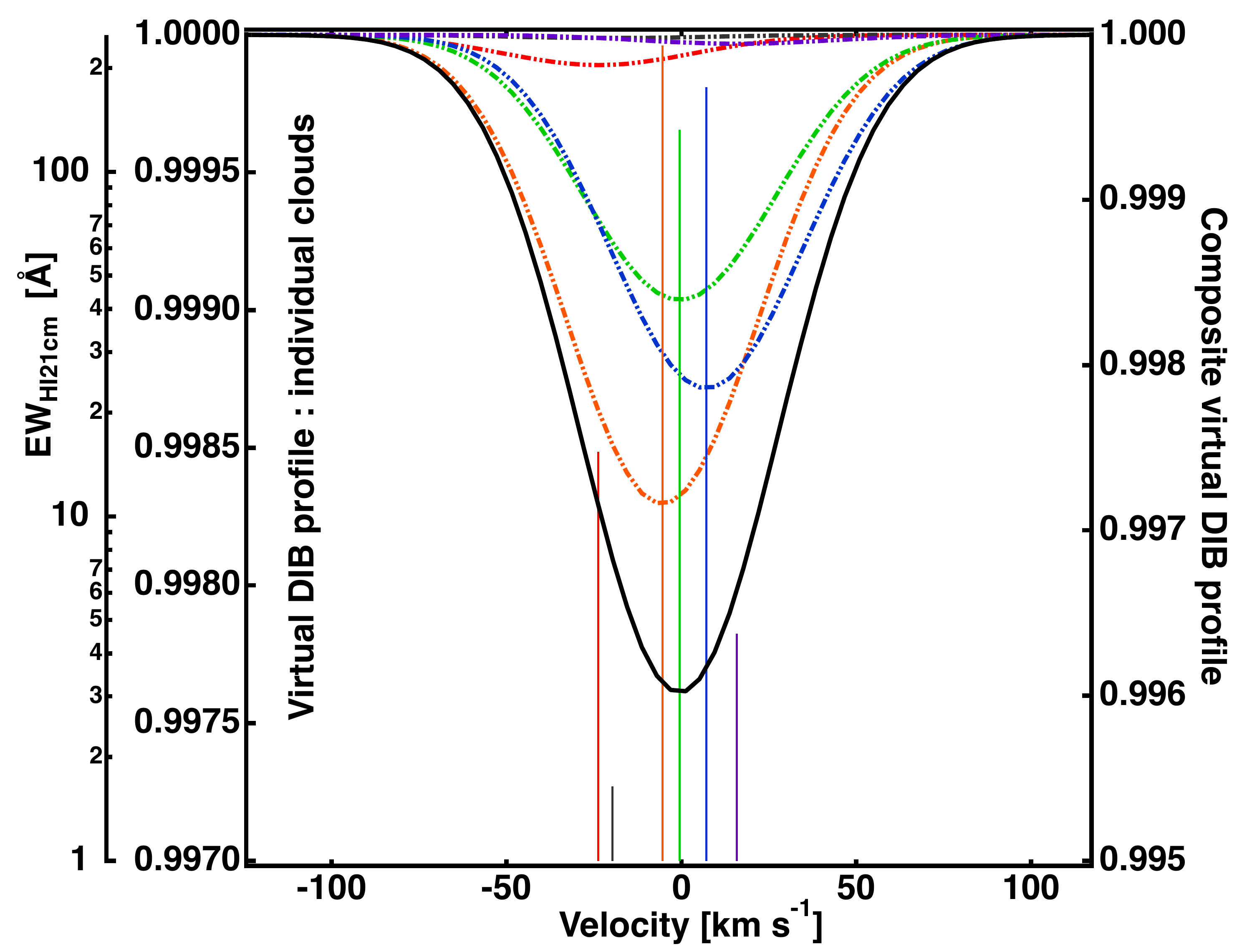}
      \caption{Example illustrating our construction of a \emph{virtual} DIB for a typical LOS with 6 HI clouds. Vertical lines in different colors mark the strength and velocity of the 6 individual HI 21cm components. The corresponding individual Gaussians for the synthetic DIBs are shown using the same colors. The function representing the total \emph{virtual} DIB through the HI disk is plotted in black. Note that for a better visualization, the scales for the total DIB and individual components are different.}
         \label{dibvirtual}
\end{figure}

\begin{figure}[ht]
\centering
\includegraphics[width=0.9\columnwidth]{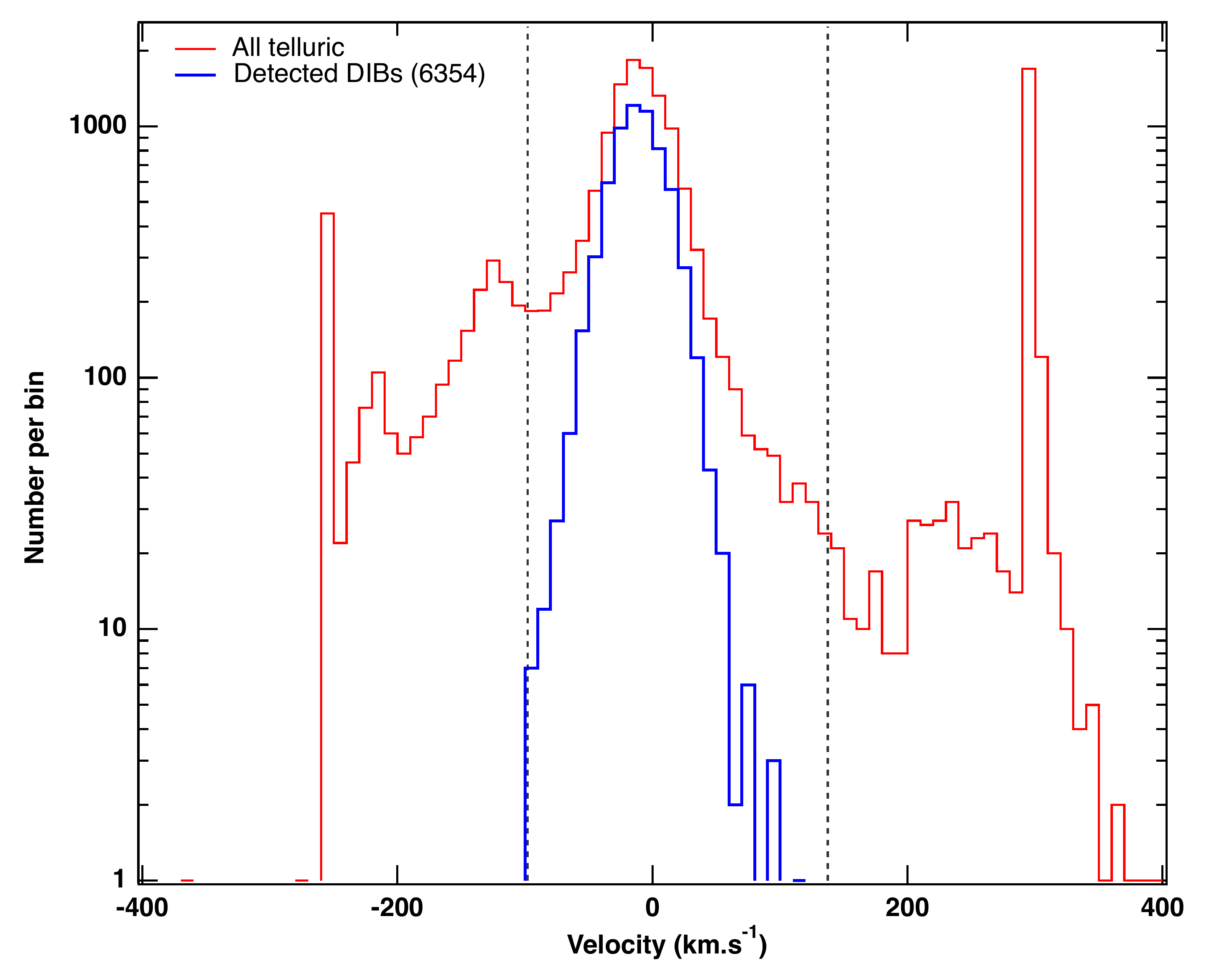}
 \caption{Histogram of the Doppler velocities for all fitted lines (red) and for 
selected DIBs only (blue). The limiting values used in the first 
selection step are shown as black vertical lines. Note the exclusion of 
all unrealistic Doppler velocities falling in the first and last bins 
and the disappearance of the secondary maximum at $\simeq$ 130 
km s$^{-1}$ that corresponds to false detections produced by a strong 
stellar line falling at this wavelength. }
 \label{velocityhist}
 \end{figure}

\begin{figure}[ht]
 \centering
 \includegraphics[width=0.9\columnwidth]{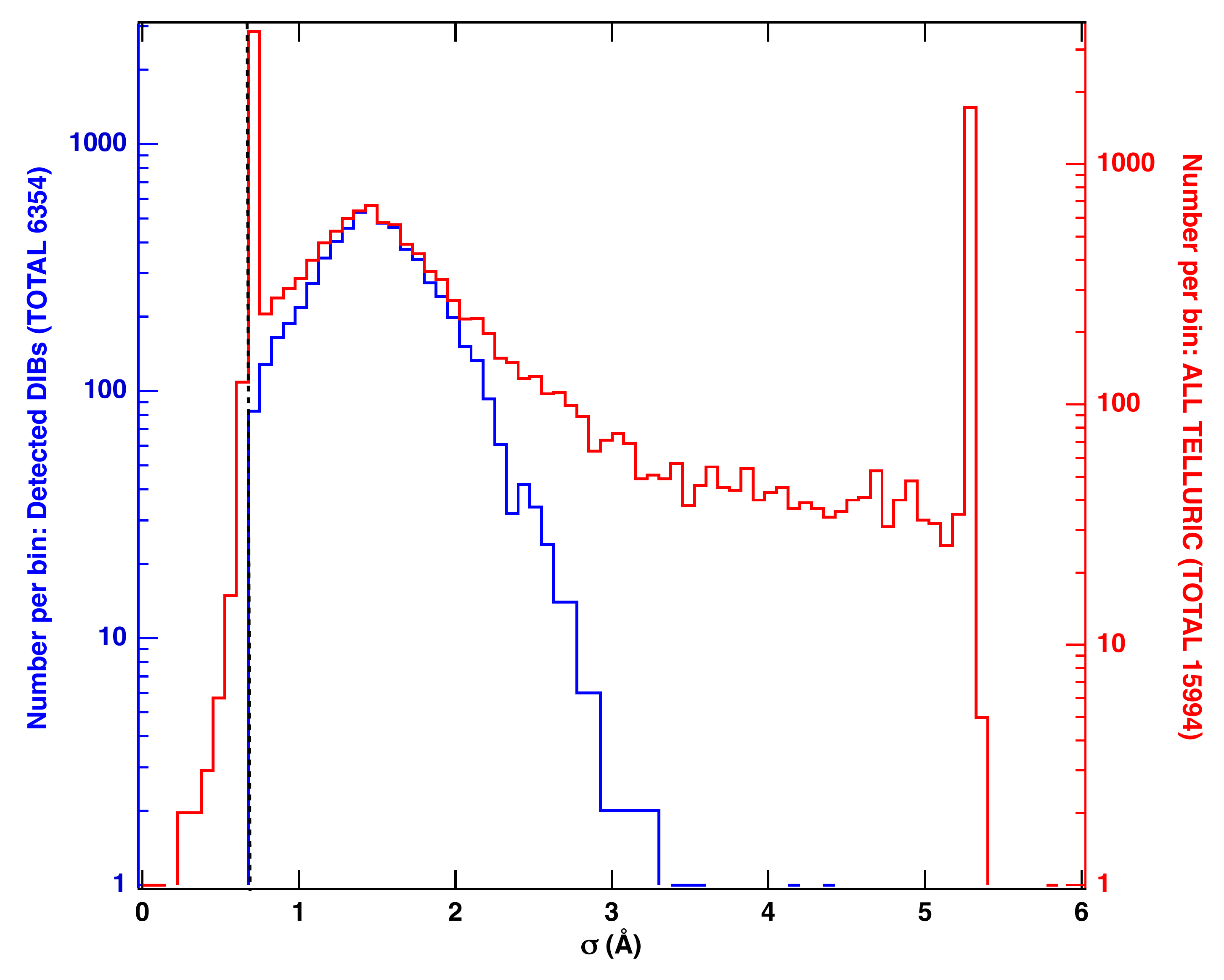}
  \caption{Same as Fig \ref{velocityhist} for the DIB width $\sigma$.  {The limiting value 0.7\AA\ used in the
selection step is shown as a black vertical line.}}
\label{sigmahist}
 \end{figure}

\begin{figure}[ht]
 \centering
 \includegraphics[width=0.9\columnwidth]{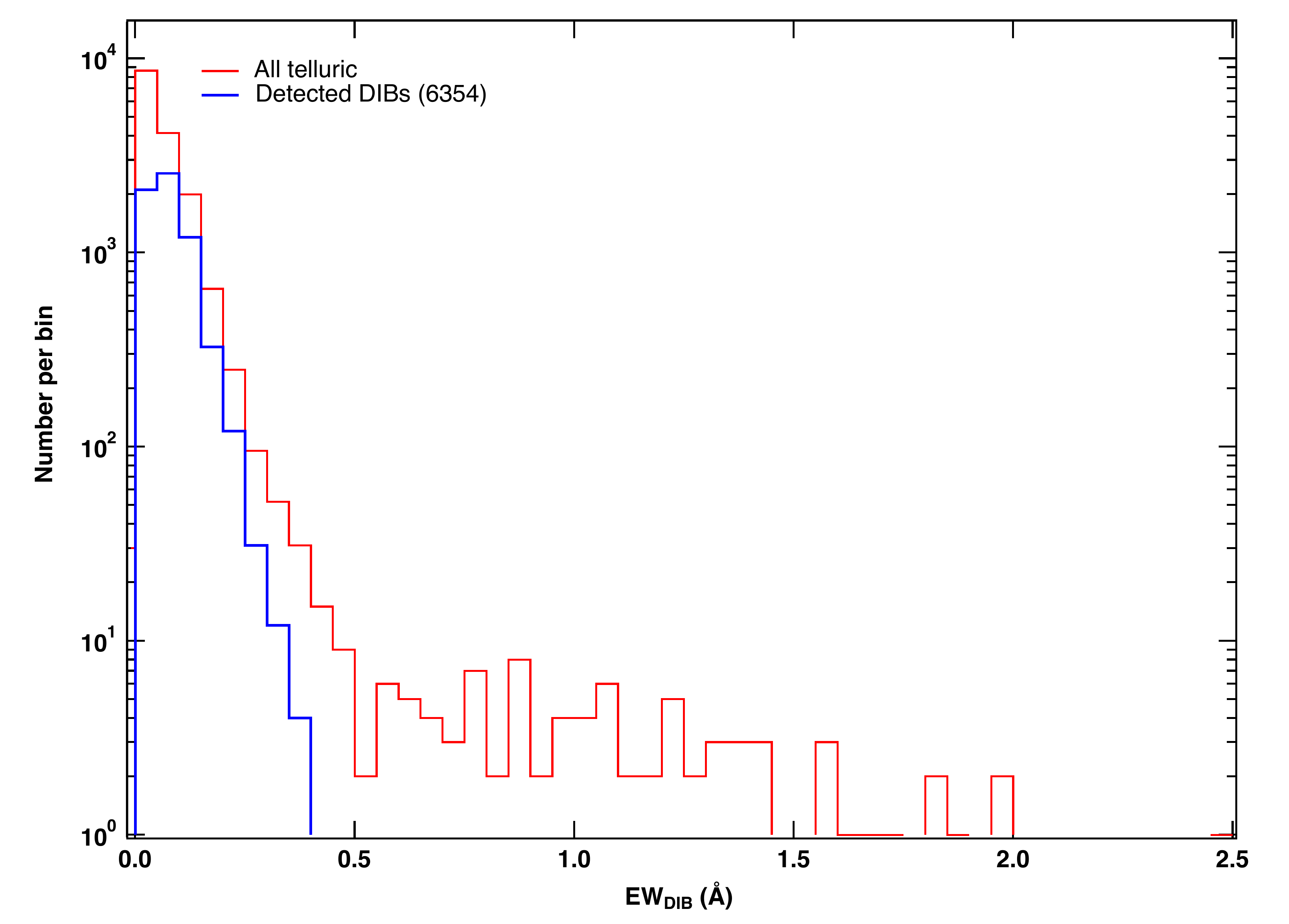}
 \caption{Same as Fig \ref{velocityhist} for the DIB equivalent width.}
  \label{ewdibhist}
 \end{figure}

\begin{figure}[ht]
   \centering
   \includegraphics[width=0.9\columnwidth,clip=]{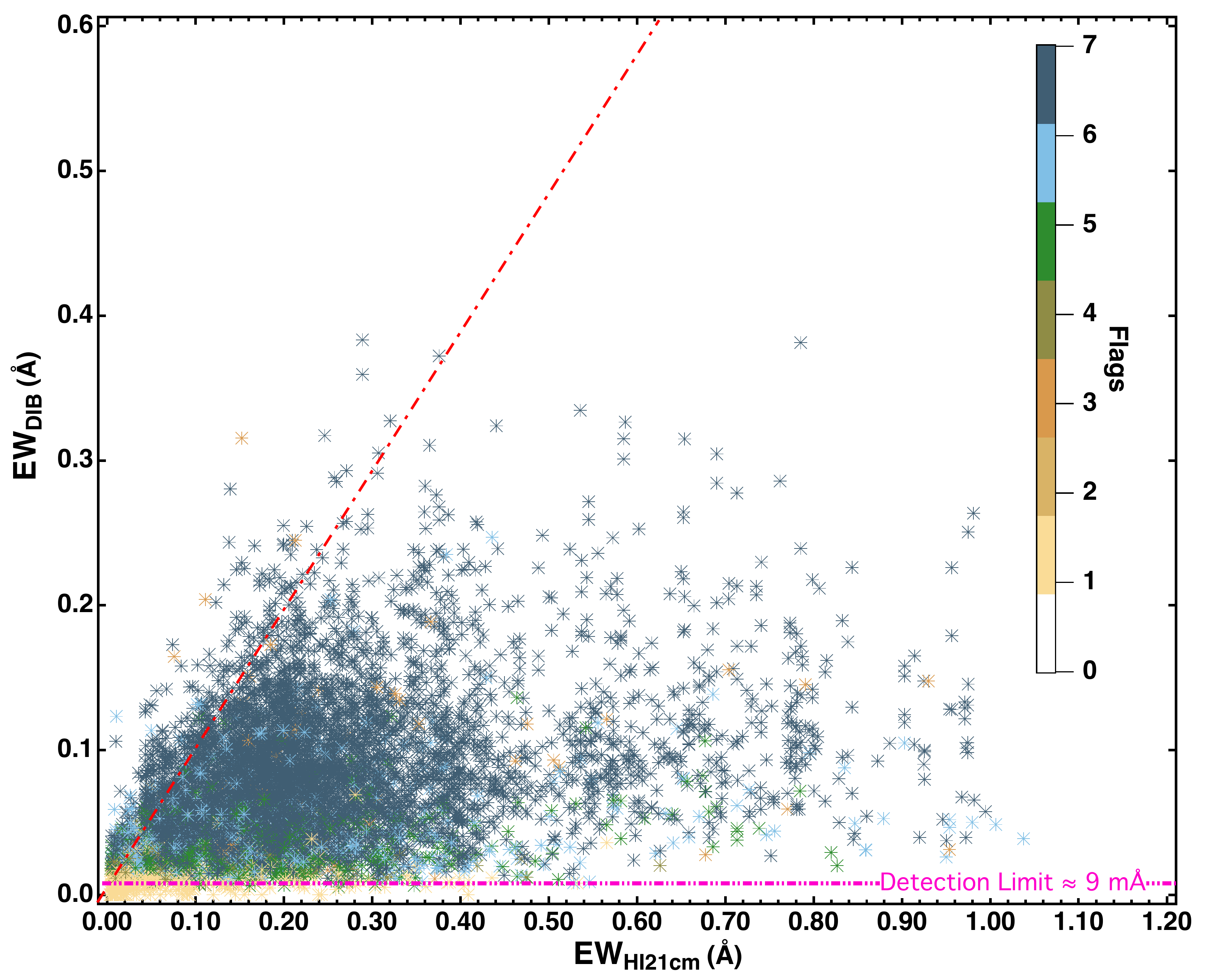}
      \caption{DIB equivalent width as a function of the \emph{virtual} DIB equivalent width that is proportional to the total HI column. The dotted red line is the one-to-one relation (see text). The color code is the same as in Fig \ref{flowchart}  with 2 additional colors indicating the special cases of Be stars (orange) and stars from field 4529 (camel) from subsection \ref {special}.  {A clear detection limit is found at around 9m\AA\, indicated by a pink dashed  line.}}
         \label{dibvsh21}
   \end{figure}

The various types of detections or upper limits resulting from the criteria shown in Figure \ref{flowchart} are illustrated in Figure
\ref{illustration}.
We assigned a flag in the catalog to each of the different categories.

\clearpage





 \begin{figure}[ht]
   \centering
   \includegraphics[width=0.9\columnwidth,clip=]{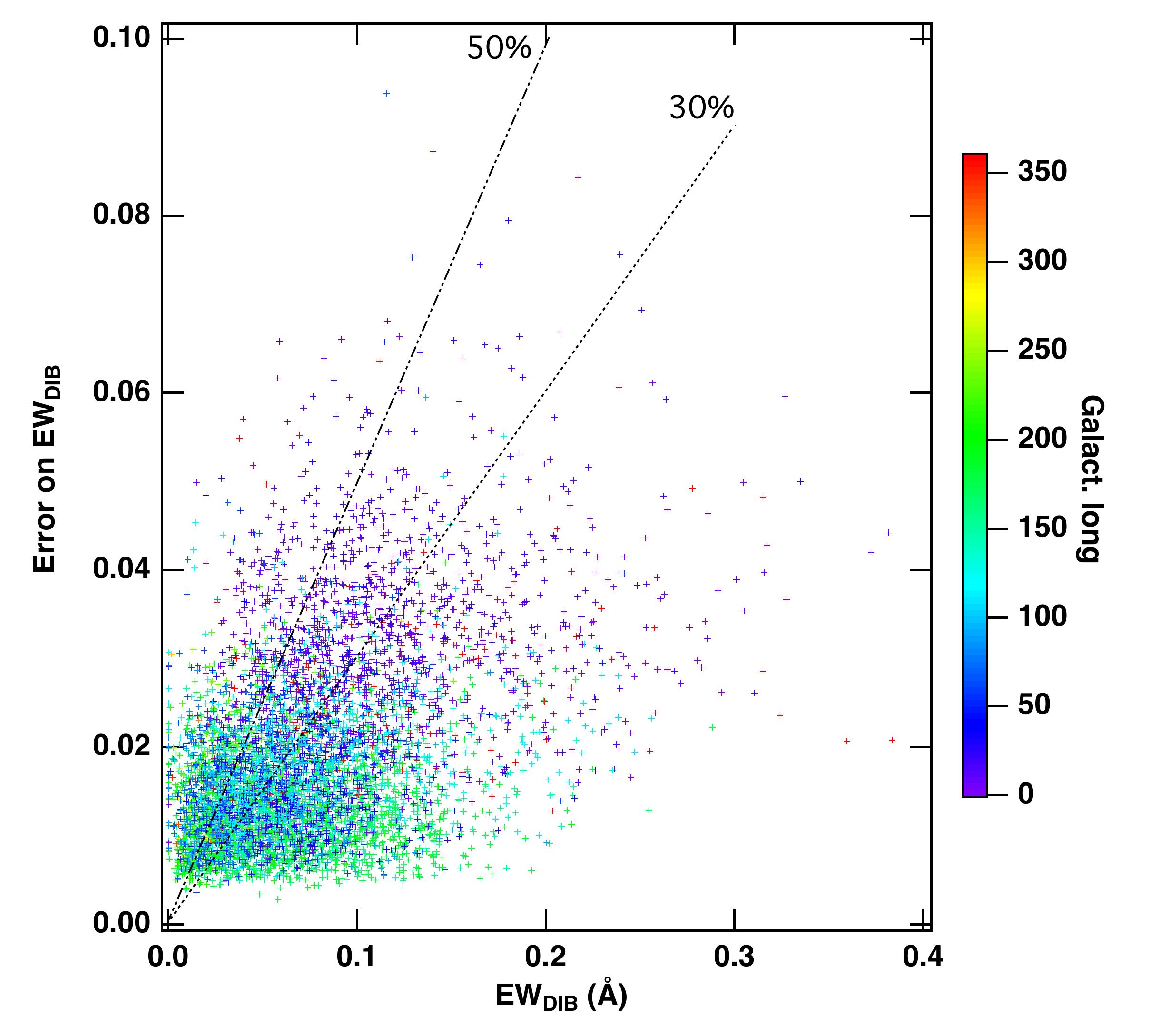}
      \caption{ {Estimated error on the DIB equivalent width EW for all the targets of the catalog. Above (resp. below) dashed lines relative errors are larger (resp. smaller) than 30 and 50 \%. The color scale refers to the Galactic longitude of the target star. Due to our method, relative errors are overestimated for targets at lower longitudes (violet, dark blue signs, see text).}}
         \label{fig9add}
   \end{figure}

\begin{figure}[ht]
   \centering
   \includegraphics[width=0.9\columnwidth,clip=]{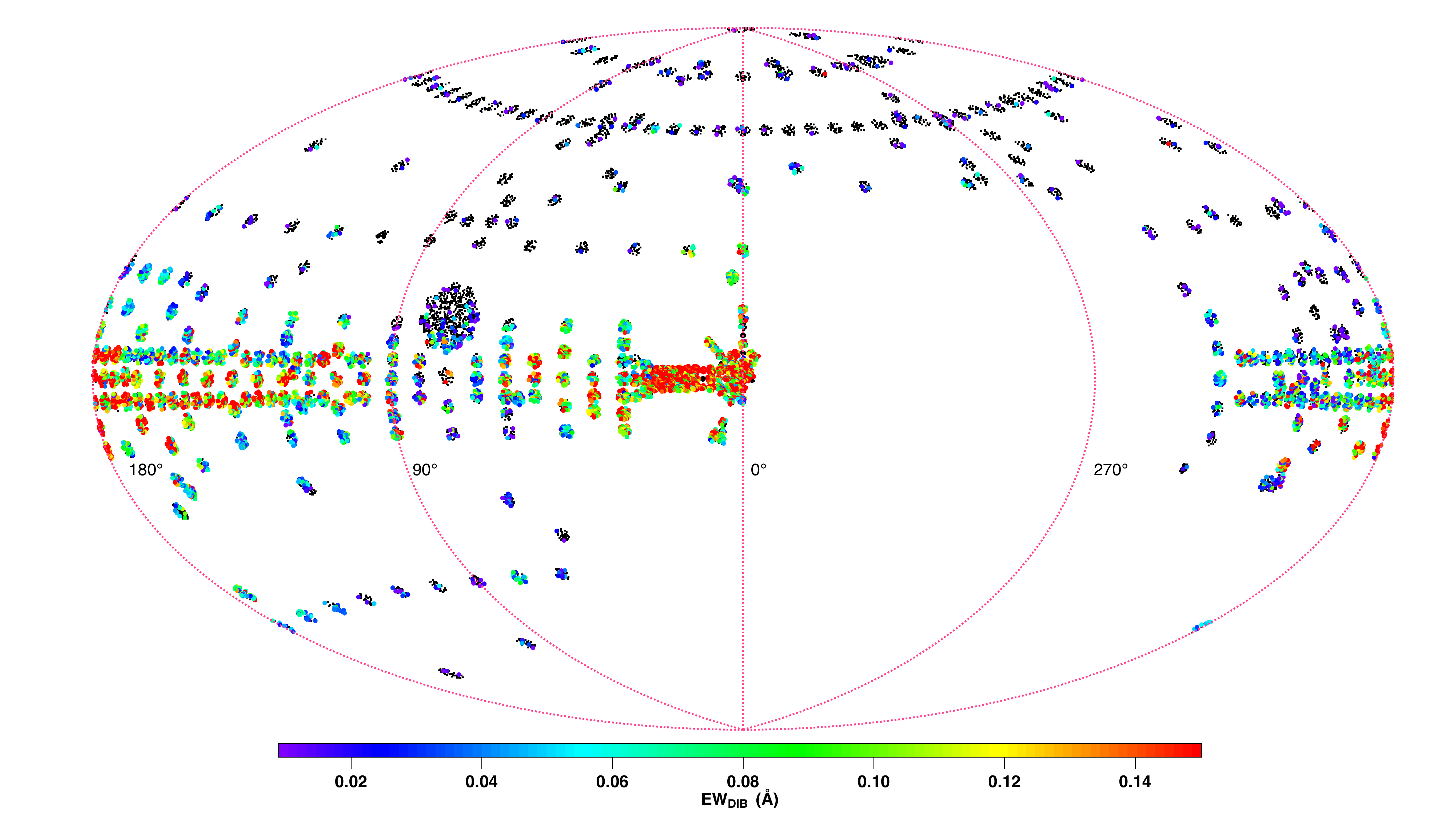}
      \caption{ {Location of all TSS targets on a sky coordinates plot, distinguishing those that are included in the catalog (colored circles, color representative of the EW) from those not included (black circles).}}
         \label{fig10add}
   \end{figure}

 \begin{figure}[ht]
   \centering
   \includegraphics[width=0.9\columnwidth,clip=]{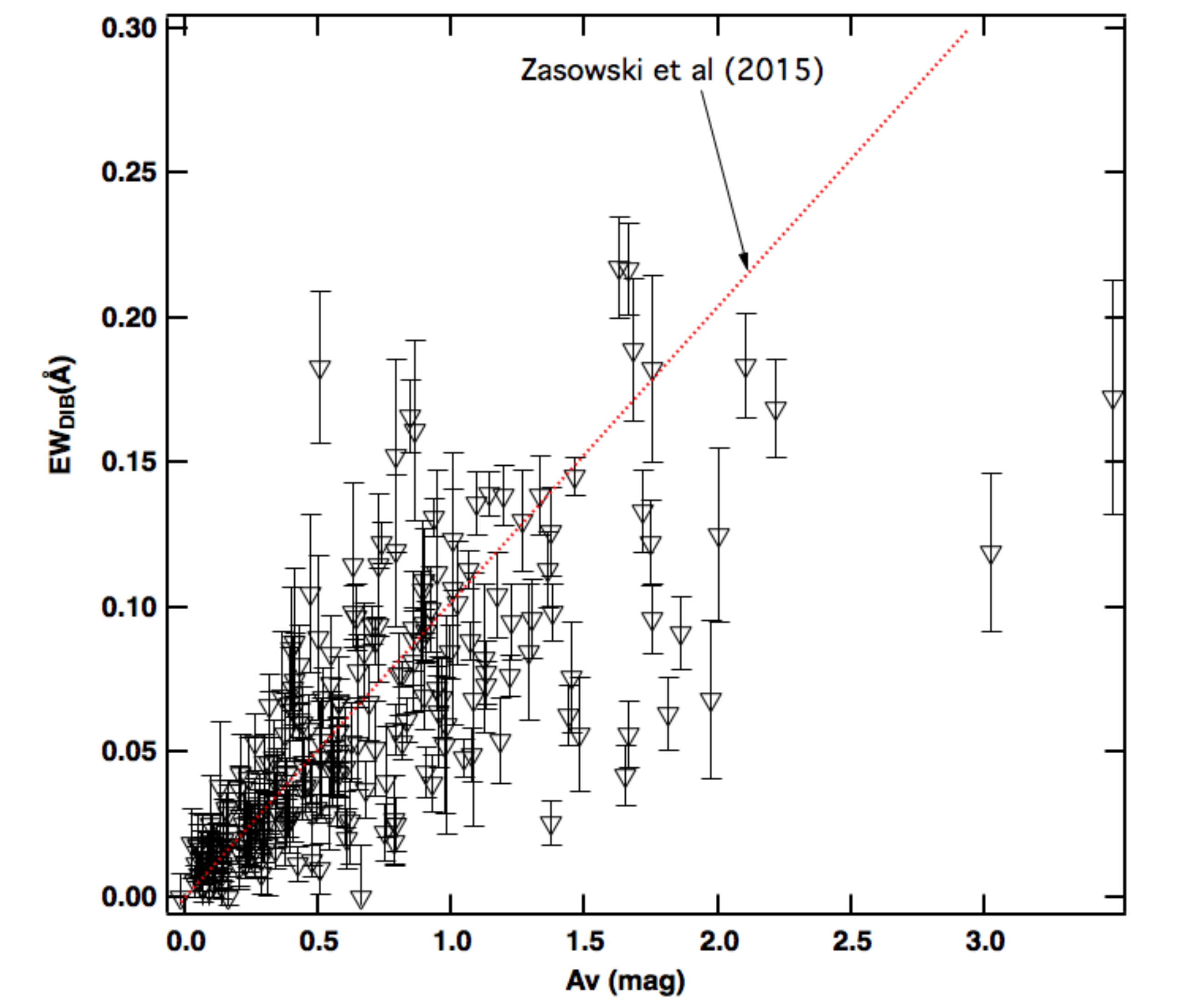}
      \caption{DIB equivalent width as a function of the color excess for the subset of 221 targets described in sect. \ref{seccatalog}. The linear relationship (red curve) is the one fitted to the whole set of APOGEE late-type stars by \citet{Zasowski15}}
         \label{dib_vs_ebv}
   \end{figure}

 \begin{figure}[ht]
   \centering
   \includegraphics[width=0.9\columnwidth,clip=]{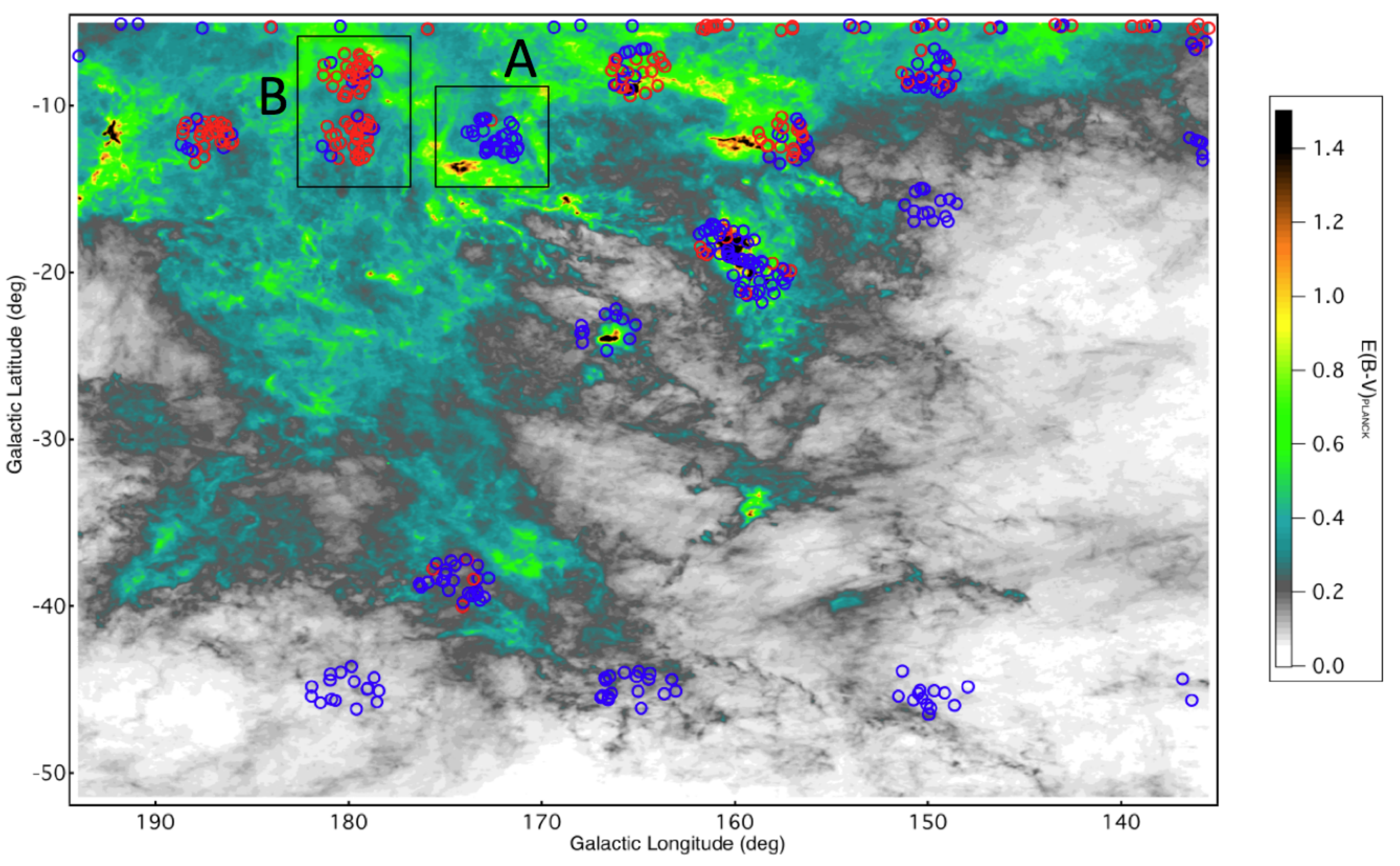}
      \caption{Planck dust optical depth in the Taurus-Perseus region with superimposed APOGEE TSS targets selected for the catalog. Blue (resp. red) markers correspond to a  DIB equivalent width lower (resp. higher) than a threshold of 50 m\AA . In regions devoid of dust (white or pale grey) DIBS are uniformly below the threshold. Towards dense clouds there is a mixed composition that reflects the location of the target (see Figure \ref{tworegions}).}
\label{plancktaurus}
   \end{figure}

\begin{figure}
\centering
 \includegraphics[width=0.45\columnwidth]{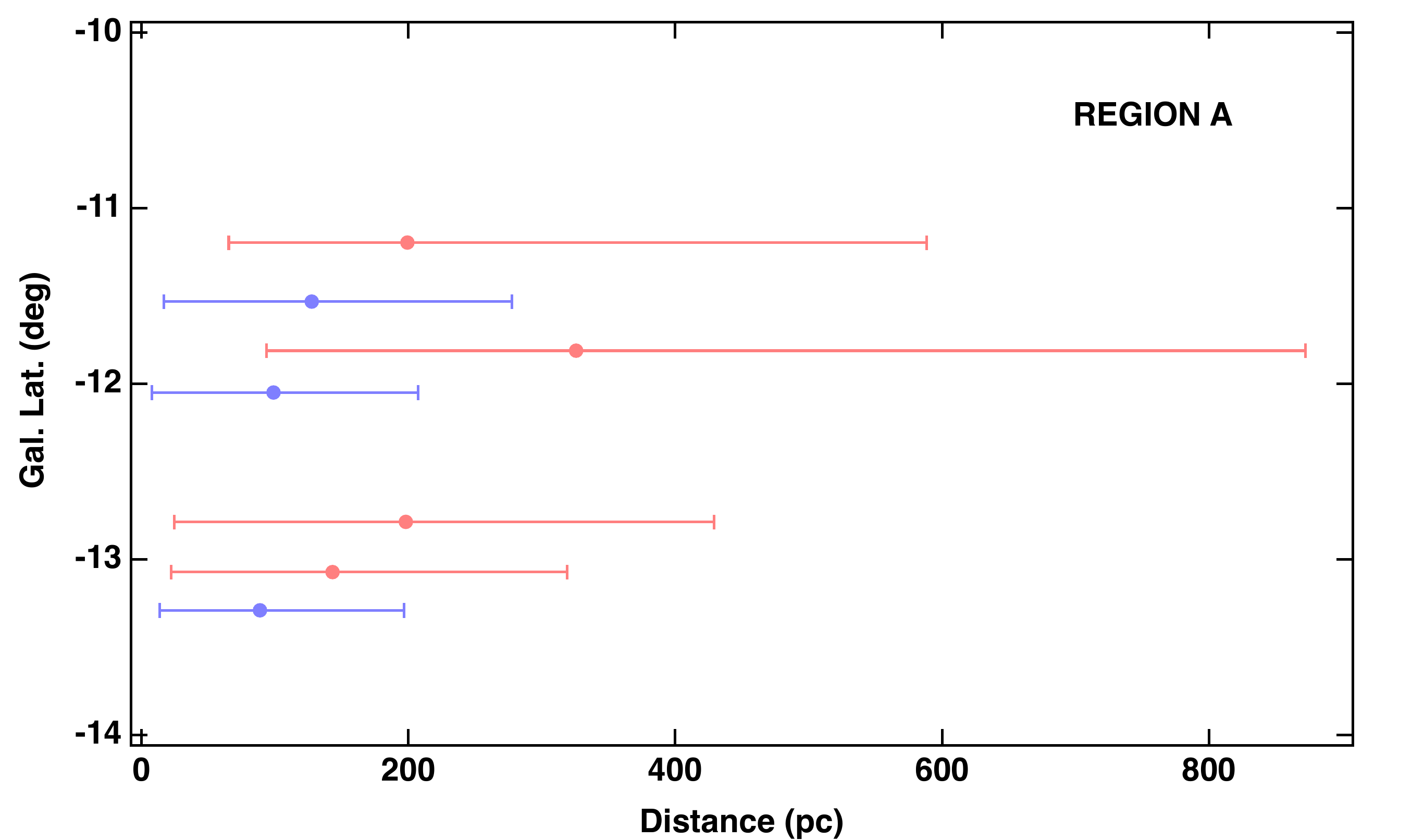}
 \includegraphics[width=0.45\columnwidth]{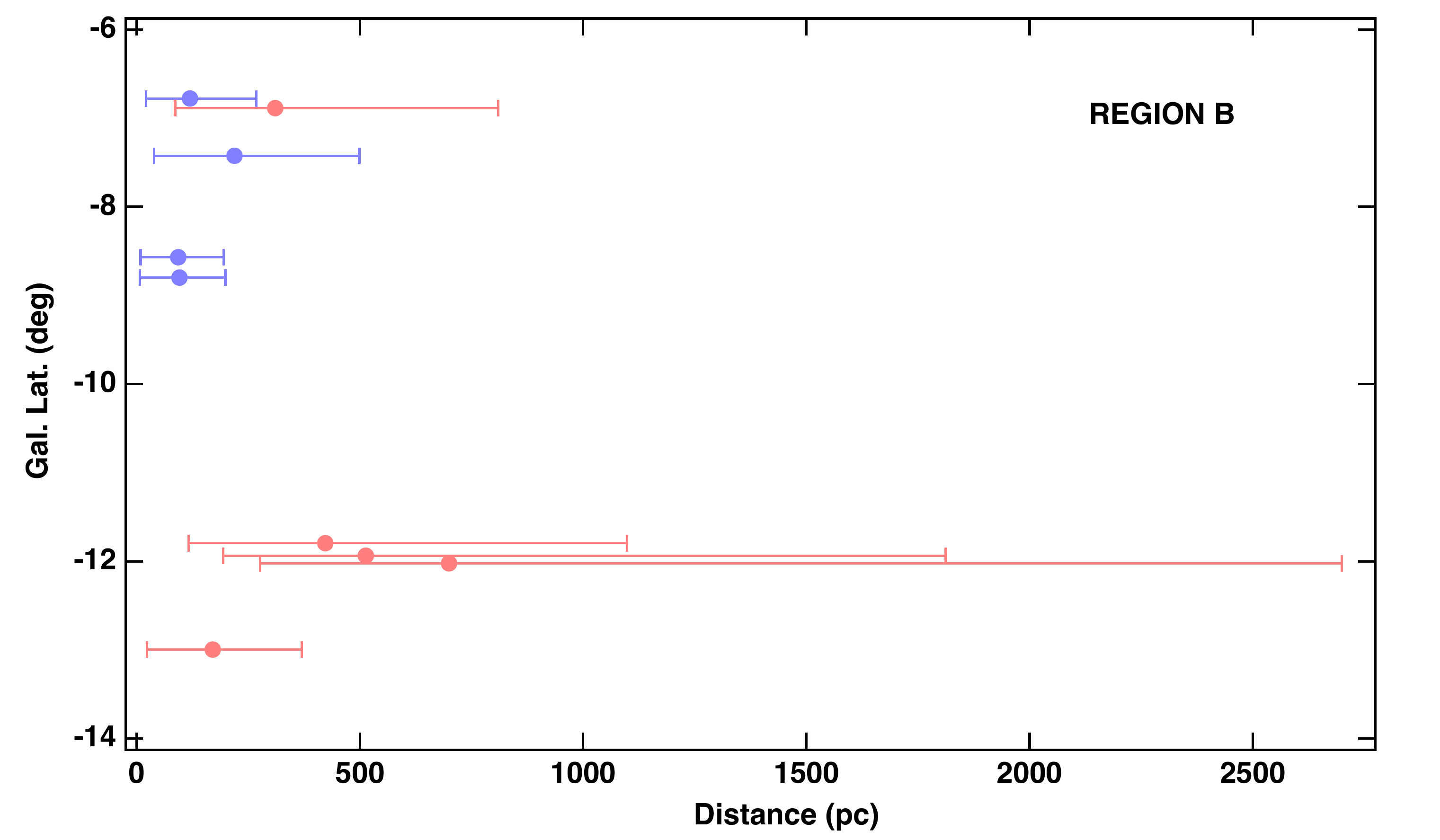}
\caption{Example of future 3D mapping based on DIBs: DIB strengths and target Hipparcos distances for the two regions A and B from Figure \ref{plancktaurus}. {  Blue (resp. red) markers correspond to a  DIB equivalent width lower (resp. higher) than a threshold of 50 m\AA.} The closest (resp. most distant) targets show consistently negligible (resp. strong) absorption. Accurate and more numerous parallaxes should allow to bracket the distance to the clouds.\label{tworegions}}
\end{figure}

\begin{figure*}[ht]
\centering
 \includegraphics[width=0.45\columnwidth,clip=]{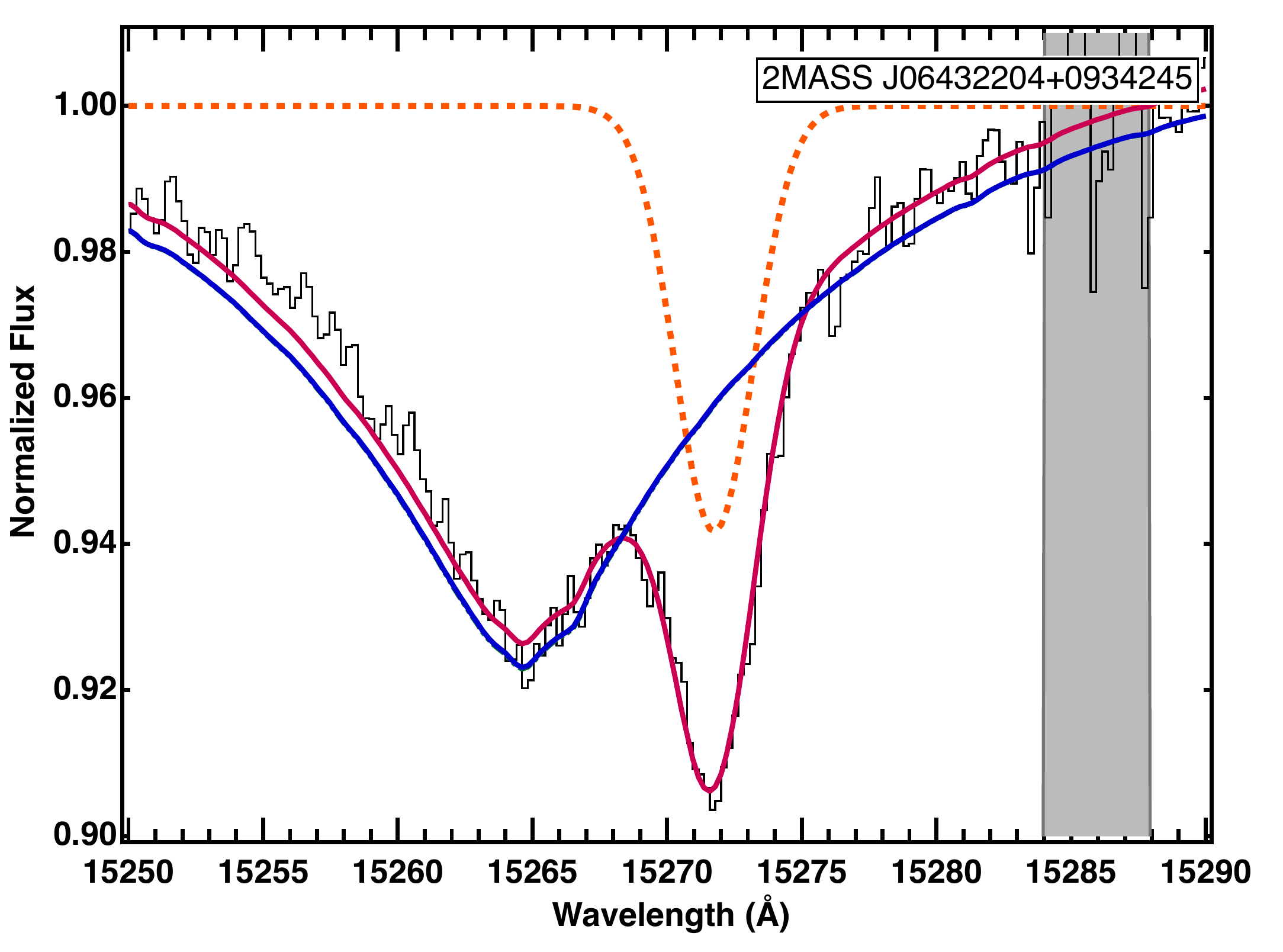}
 \includegraphics[width=0.45\columnwidth,clip=]{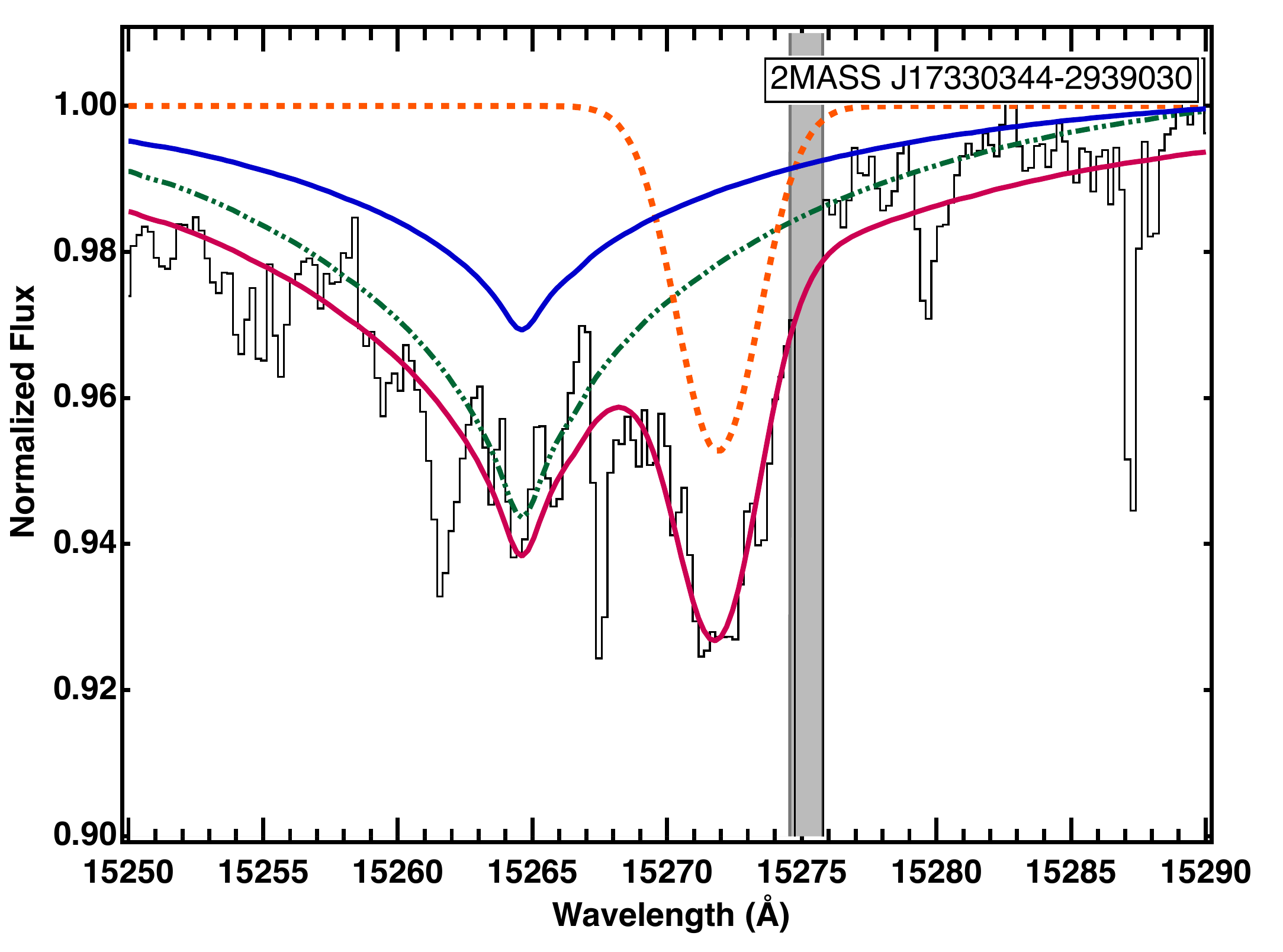}\\
 \includegraphics[width=0.45\columnwidth,clip=]{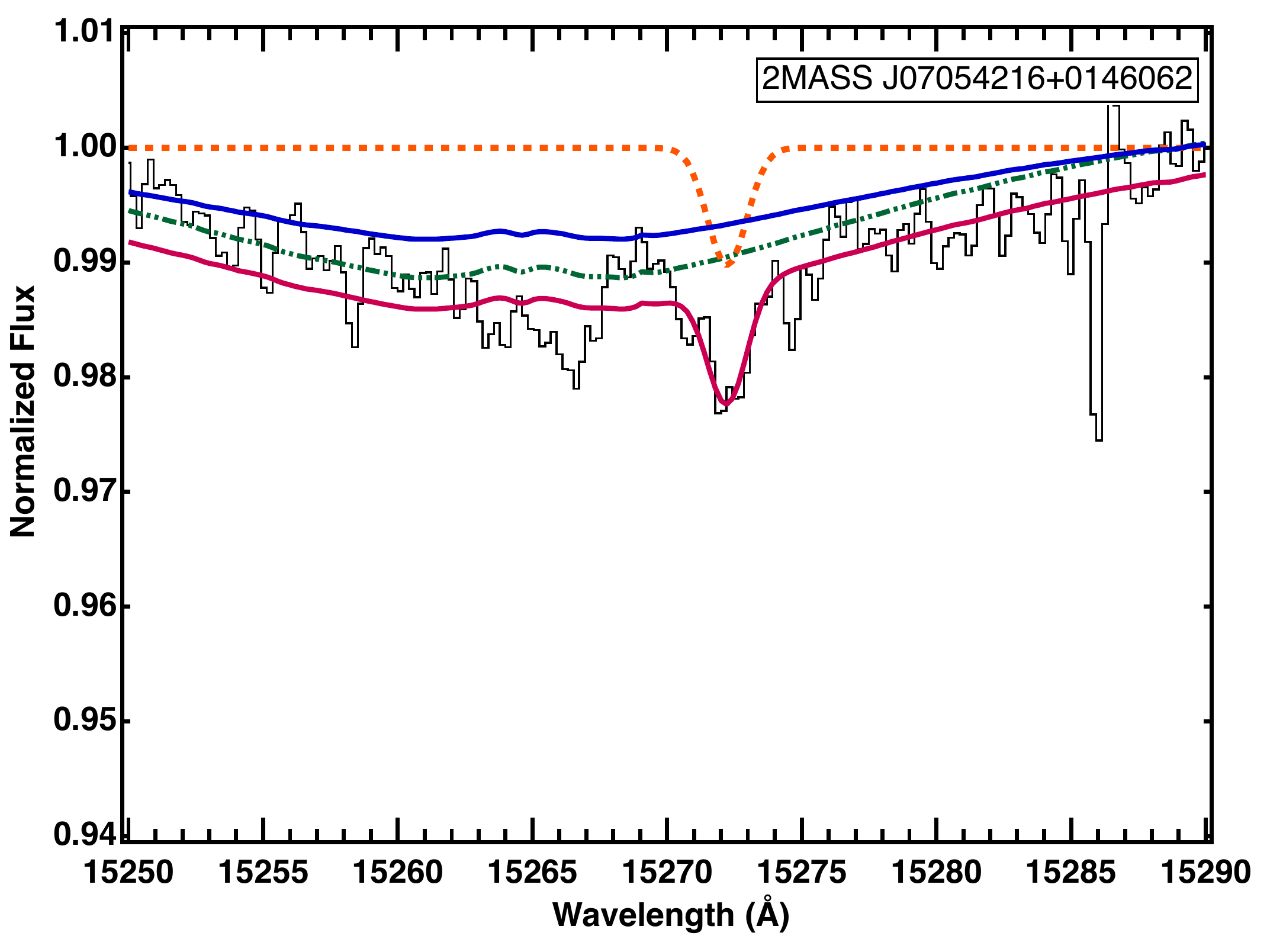}
 \includegraphics[width=0.45\columnwidth ,clip=]{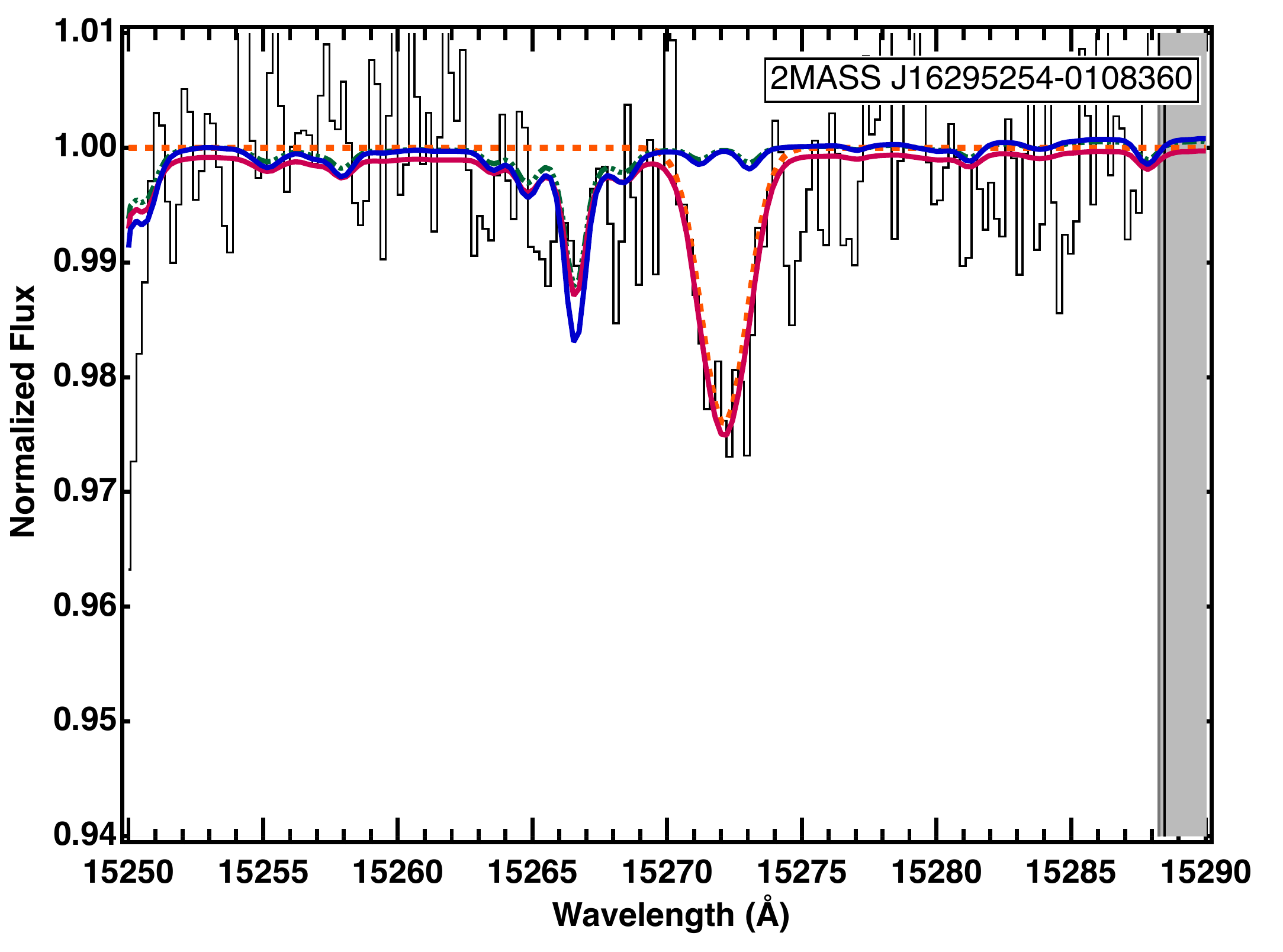}\\
 \includegraphics[width=0.45\columnwidth ,clip=]{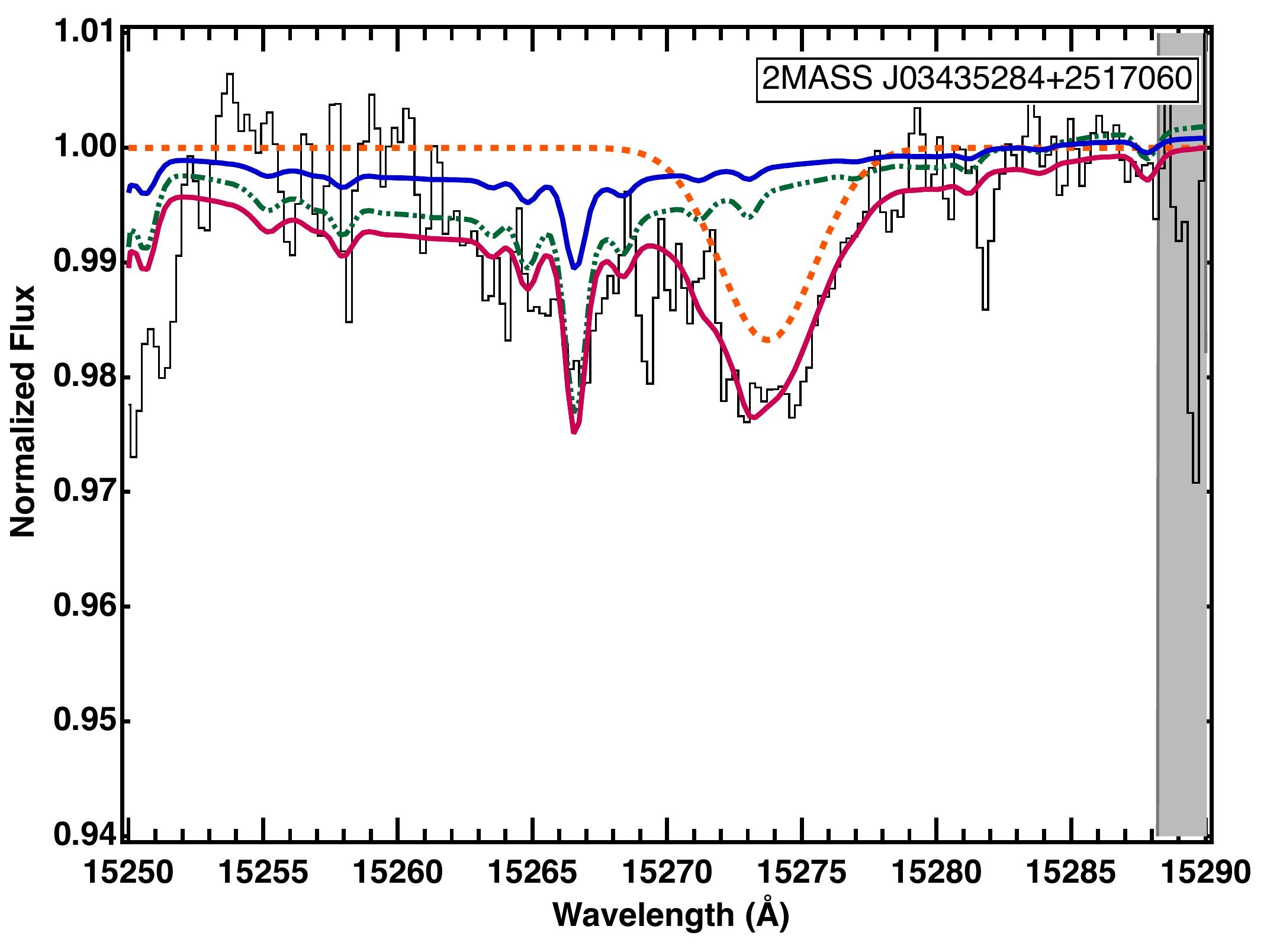}
 \includegraphics[width=0.45\columnwidth,clip=]{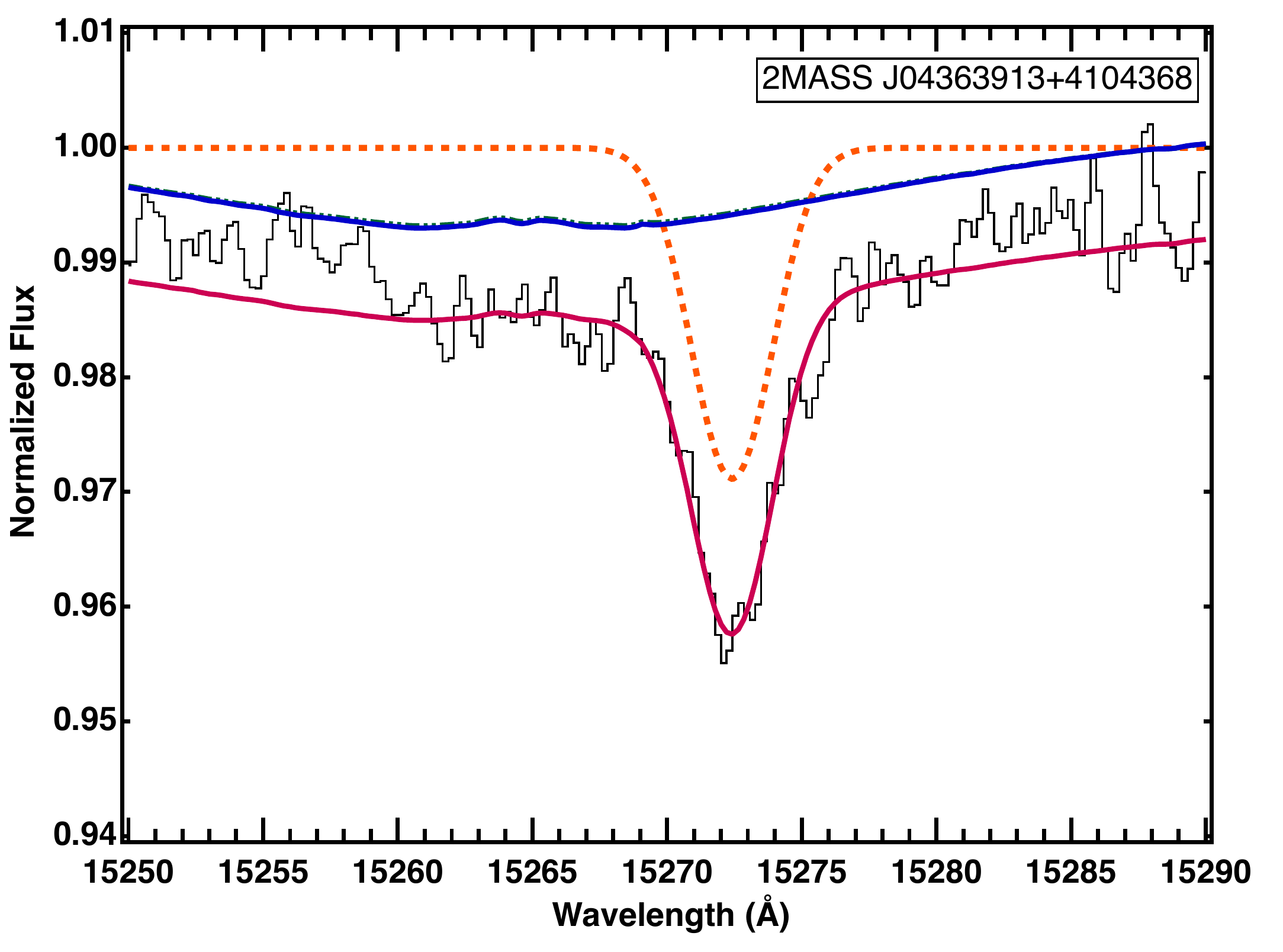}\\
\caption{\label{illustration} Illustration of the various categories of selected absorptions: (Top left) Detected-flag7; (Top right) Recovered-flag6; (Center left) Narrow-flag5; (Center right) Recovered Narrow DIBs -flag4; (Bottom left) Stars from plate 4529 -flag3; (Bottom right) Be Stars -flag2. Color and symbol code is as in Figure \ref{Figfittingmethod}.}
\end{figure*}
\newpage
\clearpage

\end{document}